\documentclass[times,sort&compress,3p,11pt,authoryear]{elsarticle}

\makeatletter
\def\ps@pprintTitle{%
 \let\@oddhead\@empty
 \let\@evenhead\@empty
 \def\@oddfoot{\centerline{\thepage}}%
 \let\@evenfoot\@oddfoot}
\makeatother

\usepackage[OT1]{fontenc}
\usepackage{amsthm,amsmath,amsfonts,amssymb,mathtools}
\usepackage{natbib,hyperref}

\usepackage{geometry}
\usepackage[tableposition=t]{caption} 

\usepackage{longtable} 
\usepackage[table]{xcolor} 
\usepackage{sectsty}
\usepackage{pdflscape} 
\usepackage{graphicx} 
\usepackage{etoolbox} 
\patchcmd{\thebibliography}{\clubpenalty4000}{\clubpenalty10000}{}{}
\patchcmd{\thebibliography}{\widowpenalty4000}{\clubpenalty10000}{}{}

\usepackage{algpseudocode,algorithm}
\usepackage{multirow}
\newcommand{\midrule}{\hline}
\newcommand{\toprule}{\hline}
\newcommand{\bottomrule}{\hline}

\newcommand{\R}{\mathbb{R}}
\newcommand{\EE}{\mathbb{E}}
\newcommand{\nvert}[0]{\, \vert \, }
\newcommand{\bb}[1]{\boldsymbol{#1}}
\newcommand{\rd}{\mathrm{d}}
\newcommand{\leqdef}{\vcentcolon=}
\newcommand{\ii}{\mathrm{i}}

\usepackage{xspace}
\newcommand{\RR}{{\fontfamily{cmss}\fontseries{sbc}\fontshape{n}\selectfont R}\xspace}

\usepackage{fancyhdr}
\fancypagestyle{mypagestyle}{%
  \fancyhf{}
  \fancyhead[LO]{A.\ Desgagn\'e, P.\ Lafaye de Micheaux and F.\ Ouimet}%
  \fancyhead[RE]{Empirical power comparison of Laplace goodness-of-fit tests}%
  \fancyfoot[C]{\thepage}%
}
\allowdisplaybreaks

\begin{document}

\begin{frontmatter}

    \title{{\bf \LARGE A comprehensive empirical power comparison of \\[-1mm] univariate goodness-of-fit tests \\[0.5mm] for the Laplace distribution}}

    \vspace{7mm}

    \author[a1]{Alain Desgagn\'e\texorpdfstring{}{)}}%
    \author[a2,a3,a4]{Pierre Lafaye de Micheaux\texorpdfstring{}{)}}%
    \author[a5,a6]{Fr\'ed\'eric Ouimet\texorpdfstring{}{)}}%

    \address[a1]{Universit\'e du Qu\'ebec \`a Montr\'eal, Montr\'eal, QC H2X 3Y7, Canada.}%
    \address[a2]{UNSW Sydney, Sydney, NSW 2052, Australia.}%
    \address[a3]{Desbrest Institute of Epidemiology and Public Health, Univ Montpellier, INSERM, Montpellier, France.}%
    \address[a4]{AMIS, Universit\'{e} Paul Val\'{e}ry Montpellier 3, Montpellier, France.}%
    \address[a5]{California Institute of Technology, Pasadena, CA 91125, USA.}%
    \address[a6]{McGill University, Montreal, QC H3A 0B9, Canada.}%

    \begin{abstract}
      In this paper we present the results from an empirical power comparison of 40 goodness-of-fit tests for the univariate Laplace distribution, carried out using Monte Carlo simulations with sample sizes $n = 20, 50, 100, 200$, significance levels $\alpha = 0.01, 0.05, 0.10$, and 400 alternatives consisting of asymmetric and symmetric light/heavy-tailed distributions taken as special cases from 11 models. In addition to the unmatched scope of our study, an interesting contribution is the proposal of an innovative design for the selection of alternatives. The 400 alternatives consist of 20 specific cases of 20 submodels drawn from the main 11 models. For each submodel, the 20 specific cases corresponded to parameter values chosen to cover the full power range. An analysis of the results leads to a recommendation of the best tests for five different groupings of the alternative distributions. A real-data example is also presented, where an appropriate test for the goodness-of-fit of the univariate Laplace distribution is applied to weekly log-returns of Amazon stock over a recent four-year period.
    \end{abstract}

    \begin{keyword}
        Laplace distribution, goodness-of-fit tests, Monte Carlo simulations, power comparison, double exponential, symmetric distributions, heavy-tailed distributions
        \MSC[2020]{Primary: 62F03}
    \end{keyword}


\end{frontmatter}

\section{Introduction}\label{sec:intro}

    The Laplace (or double-exponential) distribution is one of the most widely used probability distributions for modeling symmetric data with heavier tails than the normal distribution. Given its broad usage, testing that a sample has been drawn from a Laplace distribution is a common problem of goodness-of-fit in statistical practice. As a result, a large number of Laplace tests have been proposed in the literature. New tests appear regularly; see, e.g., \cite{Alizadeh_Noughabi_2019, Alizadeh_Noughabi_Jarrahiferiz_2019} and \cite{Desgagne_Lafaye_de_Micheaux_Ouimet_2022} for the most recent ones. Others can be developed by adapting existing strategies \cite[e.g.,][for a likelihood ratio test approach]{Pewsey_Abe_2015}.

    Comparative empirical power studies are presented in most articles introducing new Laplace goodness-of-fit tests. The number of tests and alternatives considered in such articles has been, however, rather limited. The authors usually recommend the new test they introduce, which raises issues about the choice of their alternatives. Table~\ref{tab-recommendations} gives a summary of the main references on this topic.

    \vspace{5mm}
    \noindent
    \fbox{
    \begin{minipage}{38.2em}
        \small
        This manuscript was accepted for publication in the Journal of Statistical Computation and Simulation.
        This version {\it may differ} from the published version in typographic details.
    \end{minipage}
    }

    \newpage
    \begin{table}[H]\centering
        \caption{Details of the main references on goodness-of-fit testing for the Laplace distribution.\\Abbreviations of test statistics are those used in the second column of Table~\ref{tab-selected-tests}, as well as DD \citep{Claeskens_Hjort_2004}, G \citep{Lafaye_de_Micheaux_Tran_2016}, LT \citep{Gulati_2011}, $R_n'$ \citep{Gonzalez-Estrada_Villasenor_2016}, $\sqrt{v}_1$, $v_2$ \citep{Gel_2010}, and $\tilde{V}_3$ \citep{Best_Rayner_Thas_2008}.} \label{tab-recommendations}
        \begingroup\setlength{\fboxsep}{0pt}
        \scriptsize

\begin{tabular}{m{4cm}|m{2.8cm}|m{4cm}|m{2.5cm}}
            \hline
            Publication & Tests included & Alternative distributions used & Recommendation(s) \\
            \hline
            \cite{Puig_Stephens_2000}    & AD, CvM, KS, Ku, Wa & Cauchy, logistic, normal, skew extreme value & Wa \\
                \hline
            \cite{Meintanis_2004}    & AD, CvM, KS, $\text{Me}^{(1)}_{a}$, $\text{Me}^{(2)}_{a}\quad a\in\{0.5,1,2,4,5,10\}$, Wa  & Asymmetric~Laplace, beta, Cauchy, exponential, gamma, logistic, normal, uniform & AD, $\text{Me}^{(1)}_{2}$, $\text{Me}^{(2)}_{0.5}$ \\
                \hline
            \cite{Choi_Kim_2006}    & AD, $\text{CK}_c$, $\text{CK}_e$, $\text{CK}_v$, CvM, KS, Ku, Wa & Cauchy, extreme value, inverse Gaussian, logistic, log-normal, normal, Student-$t$, uniform, Weibull, $\chi^2$  & $\text{CK}_c$, $\text{CK}_v$ \\
                \hline
            \cite{Best_Rayner_Thas_2008}    & AD, $\text{CK}_v$, DD, LK, $\text{Me}^{(1)}_{2}$, $\tilde{V}_3$, $\tilde{V}_4\equiv Ho_K$   & Beta, Cauchy, exponential, extreme value, gamma, logistic, normal, Tukey, uniform   & $\text{CK}_v$, $\tilde{V}_4$\\
                \hline
            \cite{Gel_2010}    & AD, CvM, Ge, KS, Ku, $\sqrt{v}_1$, $v_2$, Wa  & Cauchy, exponential, inverse Gaussian, logistic, normal, Tukey's contaminated normal & Ge \\
                \hline
            \cite{Gulati_2011}    & AD, CvM, KS, LT, $\text{Me}^{(1)}_{2}$, $\text{Me}^{(2)}_{0.5}$, Wa  & Beta, Cauchy, exponential, extreme value smallest element, gamma, logistic, normal, uniform  & LT \\
\hline
            \end{tabular}%
        \endgroup
    \end{table}

    In this paper we present the details of a far more extensive comparative power study. We compiled an extensive list of 40 goodness-of-fit tests for the univariate Laplace distribution: 38 identified from the literature and two new tests, the full details of which are given in \cite{Desgagne_Lafaye_de_Micheaux_Ouimet_2022}. We then performed a comparative power study of the 40 tests using Monte Carlo simulation, for sample sizes of $n = 20, 50, 100, 200$, and significance levels $\alpha = 0.01, 0.05, 0.10$. In addition to the unmatched scope of our study, another interesting contribution is the proposal of an innovative design for the selection of alternative distributions. We considered a total of 400 alternatives consisting of a variety of asymmetric and symmetric light/heavy-tailed distributions, drawn from 11 models commonly used in previous comparative studies of tests for Laplaceness. These 400 alternatives consist of 20 specific cases of 20 submodels drawn from the main 11 models. For each submodel, the 20 specific cases corresponded to parameter values chosen to cover the full power range. We considered 6 symmetric light-tailed submodels, 6 symmetric heavy-tailed submodels and 8 asymmetric submodels.

    The remainder of the paper is organized as follows. Section~\ref{sec:notation} presents the notation used throughout the paper. In Section~\ref{sec:tests}, each one of the 40 goodness-of-fit tests is presented with its original source, a brief qualitative description, its test statistic and its rejection region. In Section~\ref{sec:empirical.power.comparison}, the design of the Monte Carlo study is described and the results of the empirical power comparison are presented. An analysis of the results is given which leads to a recommendation of the best tests against five different groupings of the alternative distributions. In Section~\ref{sec:example}, an appropriate test for the goodness-of-fit of the univariate Laplace distribution is applied to weekly log-returns of Amazon stock over a recent four-year period. Conclusions are drawn in Section~\ref{sec:conclusion}. Complete tables of the empirical power results are provided in \ref{sec:power.tables.complete} and the accompanying supplementary materials document.
    Files containing the results in XLS format can be downloaded at \url{http://biostatisticien.eu/JSCS2022}.

\section{Notation}\label{sec:notation}

    The probability density function (pdf) of the $\text{Laplace}(\mu,\sigma)$ distribution is
    \begin{equation*}
        f(x \nvert \mu, \sigma) \leqdef \frac{1}{2\sigma} \exp\left(-\left|\frac{x - \mu}{\sigma}\right|\right), \quad x\in \R,
    \end{equation*}
    where $\mu\in\R$ and $\sigma>0$ denote the location and  scale parameters, respectively. Note that the mean, the median and the mode all equal $\mu$, while $\sigma$ represents both $E(|X-\mu|)$ and $\sqrt{E[(X-\mu)^2]}/\sqrt{2}$, that is the mean absolute deviation (MAD) and the standard deviation divided by $\sqrt{2}$. The corresponding cumulative distribution function (cdf) is
    \begin{equation*}
        F(x \nvert \mu, \sigma) =
        \left\{\hspace{-1mm}
        \begin{array}{ll}
            \frac{1}{2} \exp\left(-\left|\frac{x - \mu}{\sigma}\right|\right), &\mbox{if } x \leq \mu, \\[2mm]
            1 - \frac{1}{2} \exp\left(-\left|\frac{x - \mu}{\sigma}\right|\right), &\mbox{if } x \geq \mu.
        \end{array}
        \right.
    \end{equation*}
    In the standard case $(\mu,\sigma) = (0,1)$, the pdf and cdf of the Laplace distribution are, respectively,
    \begin{equation*}
        \psi(z)\leqdef \frac{1}{2} \exp(-|z|), \quad z\in \R, \quad \text{and} \quad
        \Psi(z) =
        \left\{\hspace{-1mm}
        \begin{array}{ll}
            \frac{1}{2} \exp(-|z|), &\mbox{if } z \leq 0, \\[2mm]
            1 - \frac{1}{2} \exp(-|z|), &\mbox{if } z \geq 0.
        \end{array}
        \right.
    \end{equation*}
    Following standard practice, we denote the pdf and cdf of the standard normal distribution by $\phi(\cdot)$ and $\Phi(\cdot)$, respectively.

    Consider the realizations $x_1,\dots,x_n$ of a random sample $X_1,\dots,X_n$ of size $n$. The empirical cdf is defined by
    \vspace{-1mm}
    \begin{equation*}
        \begin{aligned}
        F_n(x)
        \leqdef \frac{1}{n} \sum_{i=1}^n \bb{1}_{(-\infty,x]}(x_i)
        = \frac{j}{n}, \quad\mbox{if } x_{(j)}\le x < x_{(j+1)},~ j=0,1,\dots,n,
        \end{aligned}
    \end{equation*}
    with the convention $x_{(0)}\leqdef -\infty$ and $x_{(n+1)}\leqdef \infty$.
    The realization of the $r$-th order statistic, representing the $r$-th smallest value of this sample, is denoted by $x_{(r)}$. Also, denote
    \begin{equation*}
        z_i \leqdef \frac{x_i - \mu}{\sigma}, \qquad \hat{z}_i \leqdef \frac{x_i - \hat{\mu}_n}{\hat{\sigma}_n}, \qquad u_i \leqdef \Psi(z_i), \qquad \hat{u}_i \leqdef \Psi(\hat{z}_i),
    \end{equation*}
    and
    \begin{alignat*}{3}
            &\overline{x}_n \leqdef \frac{1}{n} \sum_{i=1}^n x_i, \qquad && \hat{\mu}_n \leqdef
                \left\{\hspace{-1mm}
                \begin{array}{ll}
                    x_{((n+1)/2)}, ~&\mbox{if } n ~\text{is odd}, \\[1mm]
                    \frac{1}{2} (x_{(n/2)} + x_{(n/2+1)}), ~&\mbox{if } n ~\text{is even},
                \end{array}
                \right. \\[2mm]
            &s_n^2 \leqdef \frac{1}{n} \sum_{i=1}^n (x_i - \overline{x}_n)^2, \qquad && \hat{\sigma}_n \leqdef \frac{1}{n} \sum_{i=1}^n |x_i - \hat{\mu}_n|.
    \end{alignat*}
    The maximum likelihood (ML) estimates of the location and scale parameters of the $\text{Laplace}(\mu,\sigma)$ distribution are given, respectively, by the sample median $\hat{\mu}_n$ and the mean absolute deviation from the median, that is $\hat{\sigma}_n$. In comparison, the method of moments (MOM) estimates are given by the sample mean $\overline{x}_n$ and the sample standard deviation divided by $\sqrt{2}$, that is $s_n/\sqrt{2}$.

\section{Goodness-of-fit tests for the Laplace distribution}\label{sec:tests}

   In this section we present the 40 Laplace goodness-of-fit tests included in our empirical power comparison. A summary of their basic details is given in Table~\ref{tab-selected-tests}. There, they are grouped into four categories, namely tests based on the empirical cumulative distribution function, on moments, on entropy, and other tests. The tests are listed alphabetically in each category based on the authors' surnames. Table~\ref{tab-selected-tests} also includes abbreviations for the tests used throughout this article, references to original sources, original abbreviations, and rejection regions.

   \def\arraystretch{1.15}
    \begin{table}[ht]\centering
    \scriptsize
        \caption{Summary of the 40 Laplace goodness-of-fit tests selected for the empirical power comparison.\\$H_0$ is rejected for: 1, large values;  2, small values; 3, small or large values.}
        \begin{tabular}{lllr}
            \hline
            Tests based on the empirical cdf & Abbreviations & Original sources and abbreviations & \hspace{-15mm}Reject $H_0$ \\
            \hline
            & & & \\ [-2.8mm]
            Anderson-Darling \eqref{eq:sec:tests.empirical.distribution.function.Anderson.Darling} & AD & \cite{Puig_Stephens_2000}, $A^2$ & 1 \\
            Cram\'er-von Mises \eqref{eq:original:sec:tests.empirical.distribution.function.Cramer.von.Mises} & CvM & \cite{Puig_Stephens_2000}, $W^2$ & 1 \\
            Kolmogorov-Smirnov \eqref{eq:original:sec:tests.empirical.distribution.function.Kolmogorov.Smirnov.D.minus.D.plus} & KS & \cite{Puig_Stephens_2000}, $\sqrt{n}D$ & 1 \\
            Kuiper \eqref{eq:original:sec:tests.empirical.distribution.function.Kuiper} & Ku & \cite{Puig_Stephens_2000}, $\sqrt{n}V$ & 1 \\
            Watson \eqref{eq:sec:tests.empirical.distribution.function.Watson} & Wa & \cite{Puig_Stephens_2000}, $U^2$ & 1 \\
            Zhang \eqref{eq:sec:tests.empirical.distribution.function.Zhang} & $\text{Z}_K$, $\text{Z}_A$, $\text{Z}_C$ & \cite{Al-Omari_Zamanzade_2017}, $\text{Z}_K$, $\text{Z}_A$, $\text{Z}_C$ & 1 \\
            \hline
            Tests based on moments & & \\
            \hline
            & & & \\ [-2.5mm]
            \begin{tabular}{l} \hspace{-2.2mm}Desgagn\'{e}-Lafaye \\ de Micheaux-Ouimet \eqref{eq:tests.moments.Desgagne.Lafaye.Ouimet} \end{tabular} & $\text{DLO}_{X}$, $\text{DLO}_{Z}$ & \cite{Desgagne_Lafaye_de_Micheaux_Ouimet_2022}, $X^{\mathrm{APD}}_{1}$, $Z(K^{\mathrm{net}}_{1})$ & 1,3 \\
            [2.3mm]
            Gel \eqref{eq:sec:tests.moments.Gel} & Ge & \cite{Gel_2010}, $K$ & 1 \\
            [0.3mm]
            \begin{tabular}{l} \hspace{-2.2mm}Gonz\'alez Estrada- \\ Villase\~nor \eqref{eq:sec:tests.moments.Gonzalez.Estrada.Villasenor.1} \end{tabular} & GV & \begin{tabular}{l} \hspace{-2.2mm}\cite{Gonzalez-Estrada_Villasenor_2016}, \\ \hspace{-2.2mm}$\sqrt{4n}(R_n-1)$ \end{tabular} & 3 \\
            [2mm]
            Hogg \eqref{eq:sec:tests.H.0.simple.Hogg} & $\text{Ho}_K, \text{Ho}_U, \text{Ho}_V, \text{Ho}_W$ & \cite{Hogg_1972}, $K$, $U$, $V$, $W$ & 3 \\
            Langholz-Kronmal \eqref{eq:sec:tests.moments.Langholz.Kronmal} & LK & \cite{Langholz_Kronmal_1991}, $K_1$ & 1 \\
            \hline
            Tests based on entropy & & \\
            \hline
            Alizadeh Noughabi (2016) \eqref{eq:original:sec:tests.empirical.density.function.Alizadeh.Noughabi} & $\text{A}_{\text{rat}}$ & \cite{Alizadeh_Noughabi_2016}, $T_n$ & 1 \\
            Alizadeh Noughabi (2019) \eqref{eq:original:sec:tests.entropy.Alizadeh.Noughabi} & $\text{A}_{\text{ent}}$ & \cite{Alizadeh_Noughabi_2019}, $DA_{mn}$ & 1 \\
            [0.5mm]
            Alizadeh Noughabi-Park (\ref{eq:entropy.estimator.H.v})--(\ref{eq:entropy.estimator.H.z}) & \begin{tabular}{l} \hspace{-2.2mm}$\text{AP}_v$, $\text{AP}_e$, $\text{AP}_y$, \\ \hspace{-2.2mm}$\text{AP}_a$, $\text{AP}_z$, $\text{AP}_y^{\scriptscriptstyle (\text{MLE})}$ \end{tabular} & \begin{tabular}{l} \hspace{-2.2mm}\cite{Alizadeh_Noughabi_Park_2016}, \\ \hspace{-2.2mm}$TV_{mn}$, $TE_{mn}$, $TY_{mn}$, $TA_{mn}$, $TZ_{mn}$, $Y_{mn}$ \end{tabular} & 1 \\
            [2.0mm]
            Choi-Kim \eqref{eq.Choi.tests} & $\text{CK}_v$, $\text{CK}_c$, $\text{CK}_e$ & \cite{Choi_Kim_2006}, $T_{m,n}^{V}$, $T_{m,n}^{C}$, $T_{m,n}^{E}$& 2 \\
            [0.5mm]
            \hline
            Other tests & & \\
            \hline
            \begin{tabular}{l} \hspace{-2.2mm}Alizadeh Noughabi- \\ Balakrishnan \eqref{eq:sec:tests.other.Alizadeh.Noughabi.Balakrishnan} \end{tabular} & \begin{tabular}{l} \hspace{-2.2mm}$\text{AB}_{\text{KL}}, \text{AB}_{\text{He}}, \text{AB}_{\text{Je}}$, \\ \hspace{-2.2mm}$\text{AB}_{\text{TV}}, \text{AB}_{\text{$\chi^2$}}$ \end{tabular} & \begin{tabular}{l} \hspace{-2.2mm}\cite{Alizadeh_Noughabi_Balakrishnan_2016}, \\ \hspace{-2.2mm}$\text{TKL}$, $\text{TH}$, $\text{TJ}$, $\text{TT}$, $\text{T}_{\chi}$ \end{tabular} & 1 \\
            [1.5mm]
            \begin{tabular}{l} \hspace{-2.2mm}Alizadeh Noughabi- \\ Jarrahiferiz \eqref{eq:original:sec:tests.other.Alizadeh.Noughabi.Jarrahiferiz} \end{tabular} & AJ & \cite{Alizadeh_Noughabi_Jarrahiferiz_2019}, $T_{mn}$ & 1 \\
            Brain-Shapiro \eqref{eq:sec:tests.other.Brain.Shapiro} & BS & \cite{Brain_Shapiro_1983, Gulati_2011}, $Z$ & 1 \\
            Kozubowski-Panorska \eqref{eq:sec:tests.other.Kozubowski.Panorska} & KP & \cite{Kozubowski_Panorska_2004}, $\widetilde{T}_n$ & 1 \\
            Meintanis \eqref{eq:original:sec:tests.empirical.characteristic.function.Meintanis} & $\text{Me}_2^{(1)}$, $\text{Me}_{0.5}^{(2)}$ & \cite{Meintanis_2004}, $T_{n,a}^{(1)}$, $T_{n,a}^{(2)}$ & 1 \\
            Subramanian-Dixit \eqref{eq:sec:tests.other.Subramanian} & SD & \cite{Subramanian_Dixit_2018}, $T$ & 3 \\
            Sz\'ekely-Rizzo \eqref{eq:original:sec:tests.other.Szekely.Rizzo} & $\text{SR}^{\star}$ & \cite{Rizzo_Haman_2016}, $E$ & 1 \\
            \hline
        \end{tabular}
        \label{tab-selected-tests}
    \end{table}

    Below, the tests are presented in four subsections, with a brief qualitative description and the relevant formulas provided for each test. To achieve uniformity of presentation, the test statistics are all expressed using the notation introduced in Section~\ref{sec:notation}. All 40 tests are location and scale invariant and are designed to test the following composite null hypothesis:
    \begin{align*}
        H_0: ~
        &X_1,X_2,\ldots,X_n \text{ is a random sample from a Laplace}(\mu,\sigma) \text{ distribution,} \\[-1mm]
        &\text{where both $\mu$ and $\sigma$ are unknown,}
    \end{align*}
    against the alternative hypothesis $H_1$ that the random sample is not from a Laplace$(\mu,\sigma)$ distribution.

    Note that in this paper we have only considered goodness-of-fit tests for data obtained by simple random sampling. We do not consider tests for different sampling procedures \cite[e.g.,][]{Mahdizadeh_2012,Al-Omari_Haq_2016,Al-Omari_Zamanzade_2017}.

    \subsection{Tests based on the empirical cumulative distribution function}\label{sec:tests.empirical.distribution.function}

    The tests in this subsection are based on the original test statistics of \cite{Anderson_Darling_1952}, \cite{Cramer_1928_first_paper,Cramer_1928_second_paper} and \cite{von_Mises_1928}, \cite{Kolmogorov_1933} and \cite{Smirnov_1948}, \cite{Kuiper_1960}, \cite{Watson_1961}, and \cite{Zhang_2002}, respectively, with $\mu$ and $\sigma$ replaced by $\hat{\mu}_n$ and $\hat{\sigma}_n$.

        $ $\\
        \noindent \textbf{Anderson-Darling test} The test statistic for this test is
            \begin{align}\label{eq:sec:tests.empirical.distribution.function.Anderson.Darling}
                \text{AD}& \leqdef n \int_{-\infty}^{\infty} \frac{\{F_n(x) - F(x\nvert \hat{\mu}_n,\hat{\sigma}_n)\}^2}{F(x\nvert \hat{\mu}_n,\hat{\sigma}_n) \{1 - F(x\nvert \hat{\mu}_n,\hat{\sigma}_n)\}} d F(x\nvert \hat{\mu}_n,\hat{\sigma}_n)\nonumber\\[1mm]
                   & = - n - \frac{1}{n} \sum_{i=1}^n \left[(2i - 1) \log(\hat{u}_{(i)}) +(2(n-i)+1)\log(1 - \hat{u}_{(i)})\right].
            \end{align}

        \noindent \textbf{Cram\'er-von Mises test} For this test, the test statistic is
            \begin{align}\label{eq:original:sec:tests.empirical.distribution.function.Cramer.von.Mises}
                \text{CvM} &\leqdef n \int_{-\infty}^{\infty} \{F_n(x) - F(x\nvert \hat{\mu}_n,\hat{\sigma}_n)\}^2 d F(x\nvert \hat{\mu}_n,\hat{\sigma}_n)\nonumber \\
                           & = \frac{1}{12n} + \sum_{i=1}^n \left\{\frac{2i - 1}{2n} - \hat{u}_{(i)}\right\}^2.
            \end{align}

        \noindent \textbf{Kolmogorov-Smirnov test} The test statistic is
            \begin{equation*}
              \text{KS} \leqdef \sqrt{n}
              \max\{D^-,D^+\},
            \end{equation*}
            where
            \begin{equation}\label{eq:original:sec:tests.empirical.distribution.function.Kolmogorov.Smirnov.D.minus.D.plus}
                \begin{aligned}
                    D^- \leqdef \max_{1 \leq i \leq n} \Big\{F(x_{(i)}\nvert \hat{\mu}_n,\hat{\sigma}_n) -\lim_{y\rightarrow x_{(i)}^-}F_n(y)\Big\}=\max_{1 \leq i \leq n} \Big\{\hat{u}_{(i)}- \frac{i-1}{n}\Big\}, \\
                    D^+ \leqdef \max_{1 \leq i \leq n} \Big\{F_n(x_{(i)}) - F(x_{(i)}\nvert \hat{\mu}_n,\hat{\sigma}_n)\Big\}=\max_{1 \leq i \leq n} \Big\{\frac{i}{n} - \hat{u}_{(i)}\Big\}.
                \end{aligned}
            \end{equation}

        \noindent \textbf{Kuiper test} The test statistic is
            \begin{equation}\label{eq:original:sec:tests.empirical.distribution.function.Kuiper}
                \text{Ku} \leqdef \sqrt{n}\left(D^- + D^+\right),
            \end{equation}
            where $D^-$ and $D^+$ are defined in \eqref{eq:original:sec:tests.empirical.distribution.function.Kolmogorov.Smirnov.D.minus.D.plus}.\\

        \noindent \textbf{Watson test} The test statistic is
            \begin{align}\label{eq:sec:tests.empirical.distribution.function.Watson}
                \text{Wa}& \leqdef n \int_{-\infty}^{\infty}
                    \left\{\hspace{-1mm}
                    \begin{array}{l}
                        F_n(x) - F(x\nvert \hat{\mu}_n,\hat{\sigma}_n) \\[1mm]
                        - \int_{-\infty}^{\infty} (F_n(x) - F(x\nvert \hat{\mu}_n,\hat{\sigma}_n)) d F(x\nvert \hat{\mu}_n,\hat{\sigma}_n)
                    \end{array}
                    \hspace{-1mm}\right\}^2 d F(x\nvert \hat{\mu}_n,\hat{\sigma}_n) \nonumber\\
               &= \text{CvM} - n\left(\frac{1}{n} \sum_{i=1}^n \hat{u}_i - \frac{1}{2}\right)^2,
            \end{align}
            where the CvM test statistic is given in \eqref{eq:original:sec:tests.empirical.distribution.function.Cramer.von.Mises}.\\

            \noindent \textbf{Zhang tests} Zhang's test statistics, respectively $Z_K$, $Z_A$ and $Z_C$, are likelihood ratio based tests. They are intended to provide more powerful analogues of the Kolmogorov-Smirnov test, the Anderson-Darling test and the Cram\'er-von Mises test, respectively. Their formulae are
            \begin{equation}\label{eq:sec:tests.empirical.distribution.function.Zhang}
            \begin{aligned}
                \text{Z}_K &\leqdef \max_{1 \leq i \leq n} \left(\Big(i - \frac{1}{2}\Big) \log\bigg[\frac{i - \frac{1}{2}}{n \hat{u}_{(i)}}\bigg] + \Big(n - i + \frac{1}{2}\Big) \log \bigg[\frac{n - i + \frac{1}{2}}{n (1 - \hat{u}_{(i)})}\bigg]\right),  \\[1mm]
                \text{Z}_A &\leqdef - \sum_{i=1}^n \bigg[\frac{\log(\hat{u}_{(i)})}{n - i + \frac{1}{2}} + \frac{\log(1 - \hat{u}_{(i)})}{i - \frac{1}{2}}\bigg], \\
                \text{Z}_C &\leqdef \sum_{i=1}^n \left[\log\left(\frac{\hat{u}_{(i)}^{-1} - 1}{(n - \frac{1}{2}) / (i - \frac{3}{4}) - 1}\right)\right]^2.
            \end{aligned}
            \end{equation}

    \subsection{Moment-based tests}\label{sec:tests.moments}

        \noindent  \textbf{Desgagn\'{e}-Lafaye de Micheaux-Ouimet tests} These two tests are derived from Rao's score test applied to the asymmetric power distribution (APD), a generalization of the (symmetric) exponential power distribution (EPD); see \cite{Desgagne_Lafaye_2018} and \cite{Desgagne_Lafaye_de_Micheaux_Ouimet_2022}. Their test statistics are
            \begin{equation}\label{eq:tests.moments.Desgagne.Lafaye.Ouimet}
                \text{DLO}_{X} \leqdef Z^2(S_{1})+Z^2(K^{\mathrm{net}}_{1}), \quad \text{DLO}_{Z} \leqdef Z(K^{\mathrm{net}}_{1}),
            \end{equation}
            where
            \vspace{-2.5mm}
            \begin{align*}
                Z(S_{1})&\leqdef\frac{n^{1/2} S_{1}(\mathbf{X}_n)}{\big(1 - 1.856/n^{1.06}\big)^{1/2}},\\
                Z(K^{\mathrm{net}}_{1})&\leqdef \frac{n^{1/2}\Big(\big(K^{\mathrm{net}}_{1}(\mathbf{X}_n)\big)^{1/4} - (1-\gamma)^{1/4}\left(1 -
                0.422/n^{1.01}\right)\Big)}{\Big[\frac{1}{16}(1-\gamma)^{-3/2} (\pi^2/3-3)\left(1 -
                1.950/n^{0.92}+ 39.349/n^{2.3}\right)\Big]^{1/2}},
            \end{align*}
            for an even sample size $n$, and
            \begin{align*}
               Z(S_{1})&\leqdef\frac{n^{1/2} S_{1}(\mathbf{X}_n)}{\big(1 - 0.281/n^{1.03}\big)^{1/2}},\\
               Z(K^{\mathrm{net}}_{1})&\leqdef \frac{n^{1/2}\Big(\big(K^{\mathrm{net}}_{1}(\mathbf{X}_n)\big)^{1/4} - (1-\gamma)^{1/4}\left(1 -
               0.198/n^{0.86}\right)\Big)}{\Big[\frac{1}{16}(1-\gamma)^{-3/2} (\pi^2/3-3)\left(1 -
               3.827/n^{1.04}\right)\Big]^{1/2}},
            \end{align*}
            for an odd sample size $n$, where $\gamma\leqdef 0.577215665\ldots$ denotes the Euler-Mascheroni constant. The first-power skewness -- a standardized difference between the sample mean and the sample median -- and the first-power kurtosis are given by
            \begin{equation}
                \begin{aligned}
                    S_{1}(\mathbf{X}_n)
                    &\leqdef \frac{1}{n} \sum_{i=1}^n |\hat{z}_i|\,\mathrm{sign}(\hat{z}_i)=\frac{1}{n} \sum_{i=1}^n \hat{z}_i=\hat{\sigma}_n^{-1}\big(\bar{x}_n-\hat{\mu}_n\big), \\
                    K_{1}(\mathbf{X}_n)
                    &\leqdef \frac{1}{n} \sum_{i=1}^n  |\hat{z}_i|\log|\hat{z}_i|,
                \end{aligned}
            \end{equation}
            while the first-power net kurtosis is given by
            \begin{equation}
                K^{\mathrm{net}}_{1}(\mathbf{X}_n)\leqdef \max\big(0,K_{1}(\mathbf{X}_n)-(1/2)S^2_{1}(\mathbf{X}_n)\big).
            \end{equation}
            \noindent{\it Note 1:} For $n\ge20$, the null distributions of $Z(S_{1})$ and $Z(K^{\mathrm{net}}_{1})$ can be approximated with high numerical precision by a $N(0,1)$. Moreover, $Z(S_{1})$ and $Z(K^{\mathrm{net}}_{1})$ are virtually independent. One can thus consider, for practical purposes, that $\text{DLO}_{X} \sim\chi_2^2$ and $\text{DLO}_{Z} \sim N(0,1)$. As a result, critical values and $p$-values can be estimated with high precision without the need for simulation. Furthermore, a positive (negative) value of $Z(S_{1})$ is associated with a right-skewed (left-skewed) distribution and a positive (negative) value of $\text{DLO}_{Z} \leqdef Z(K^{\mathrm{net}}_{1})$ is associated with heavy (light) tails.

            \vspace{3mm}\noindent{\it Note 2:} The $Z(S_{1})$ statistic can also be used as a Laplace test designed to detect asymmetry. However, it is practically equivalent to the KP test; see Note~2 below Equation~\eqref{eq:sec:tests.other.Kozubowski.Panorska}. To avoid duplication, only the KP test was considered in our empirical power comparison.\\

        \noindent \textbf{Gel test} This test is to the Laplace distribution what the Jarque-Bera test is to the normal distribution. The test statistic is
            \begin{equation}\label{eq:sec:tests.moments.Gel}
                \text{Ge} \leqdef \frac{n}{C_1} (\sqrt{v_1})^2 + \frac{n}{C_2} (v_2 - 6)^2,
            \end{equation}
            where
            \begin{equation*}
                \sqrt{v_1} \leqdef \frac{1}{n} \sum_{i=1}^n \left(\frac{x_i - \overline{x}_n}{\sqrt{2}\hat{\sigma}_n}\right)^3, \quad v_2 \leqdef \frac{1}{n} \sum_{i=1}^n \left(\frac{x_i - \overline{x}_n}{\sqrt{2}\hat{\sigma}_n}\right)^4,
            \end{equation*}
            and $C_1 \leqdef 60$, $C_2 \leqdef 1200$ as recommended by the author for small to moderate sample sizes.\\

        \noindent \textbf{Gonz\'alez Estrada-Villase\~nor test} This test is based on a ratio of estimators of the scale parameter $\sigma$. Its numerator is the MOM estimator and its denominator is the mean absolute deviation around the sample mean. The test statistic is
            \begin{align}\label{eq:sec:tests.moments.Gonzalez.Estrada.Villasenor.1}
                \text{GV} \leqdef \sqrt{4n} \Bigg(\frac{s_n/\sqrt{2}}{\frac{1}{n} \sum_{i=1}^n |x_i - \overline{x}_n|} - 1\Bigg).
            \end{align}
            \noindent{\it Note:} The test statistic $\sqrt{4n} (\hat{\sigma}_n^{-1}s_n/\sqrt{2} - 1)$ was also proposed by the authors, but it is equivalent to the $\text{Ho}_U$ test in \eqref{eq:sec:tests.H.0.simple.Hogg}.\\

        \noindent \textbf{Hogg tests} These four tests are based on different ratios of dispersion measurements. $\text{Ho}_K$ is the classical sample kurtosis, $\text{Ho}_U$ is proportional to the ratio of the MOM to the ML estimators of $\sigma$, and $\text{Ho}_V$ and $\text{Ho}_W$ are proportional to the ratio of the range to the ML and MOM estimators of $\sigma$ respectively. Specifically,
            \begin{equation}
                \text{Ho}_K \leqdef \frac{1}{n} \sum_{i=1}^n \left(\frac{x_i - \overline{x}_n}{s_n}\right)^4, \quad
                \text{Ho}_U \leqdef \frac{s_n}{\hat{\sigma}_n}, \quad
                \text{Ho}_V \leqdef \frac{x_{(n)} - x_{(1)}}{2 \hat{\sigma}_n}, \quad
                \text{Ho}_W \leqdef \frac{x_{(n)} - x_{(1)}}{2 s_n}. \label{eq:sec:tests.H.0.simple.Hogg}
            \end{equation}
            \noindent{\it Note:}  The test based on $\text{Ho}_K$ is equivalent to a test proposed by \cite{Rayner_Best_1989} and \cite{Best_Rayner_Thas_2008}, based on the Gram-Schmidt orthogonalization of the polynomials $\{z\mapsto z^j\}_{j=0}^{\infty}$ in $L^2(\psi(z) dz)$.\\

        \noindent \textbf{Langholz-Kronmal test} This test is based on the first trigonometric moments. More specifically,
            \begin{equation}\label{eq:sec:tests.moments.Langholz.Kronmal}
                \text{LK} \leqdef V^{-1} \cdot 2 n \left(W_1^2+W_2^2\right),
            \end{equation}
            where
            \begin{equation*}
                W_1  \leqdef \frac{1}{n} \sum_{i=1}^n \cos\left(2\pi \Psi\left(\frac{x_i-\overline{x}_n}{s_n/\sqrt{2}}\right)\right)\quad\text{and}\quad W_2 \leqdef \frac{1}{n} \sum_{i=1}^n \sin\left(2\pi \Psi\left(\frac{x_i-\overline{x}_n}{s_n/\sqrt{2}}\right)\right)
            \end{equation*}
            and the value of the constant $V^{-1}$ is set to 0.928 for the asymptotic variance of the test statistic to be 4 (since the asymptotic distribution is $\chi_2^2$). 

            \vspace{3mm}\noindent{\it Note:} A positive (negative) value of $W_1$ indicates a light-tailed (heavy-tailed) distribution and a positive (negative) value of $W_2$ indicates a right-skewed (left-skewed) distribution.

    \subsection{Entropy based tests}\label{sec:tests.entropy}

        \noindent \textbf{Alizadeh Noughabi (2016) test} The author proposed a density-based empirical likelihood ratio (DBELR) test. It is a modified version of the likelihood ratio test of \cite{Vexler_Gurevich_2010}, where the density function under $H_1$ is replaced by an entropy-based estimate. The test statistic is
            \begin{equation}\label{eq:original:sec:tests.empirical.density.function.Alizadeh.Noughabi}
                \text{A}_{\text{rat}} = \min_{1 \leq m < (n^{\delta}\wedge \frac{n}{2})} \prod_{j=1}^n \frac{2m}{n(x_{((j+m) \wedge n)} - x_{((j-m) \vee 1)})f(x_j \nvert \hat{\mu}_n,\hat{\sigma}_n)},
            \end{equation}
            where $\delta\in (0,1)$ is set to 0.5 as \cite{Alizadeh_Noughabi_2016} did in his power comparisons and the window size $m$ is a positive integer. Note that we have added the upper bound $m < n/2$ to be consistent with other entropy estimators.\\

        \noindent \textbf{Alizadeh Noughabi (2019) test} The test statistic is minus the entropy estimator of \cite{Vasicek_1976} for the sample $\hat{u}_1, \hat{u}_2, \dots, \hat{u}_n$, namely
            \begin{equation}\label{eq:original:sec:tests.entropy.Alizadeh.Noughabi}
                \text{A}_{\text{ent}} = - \frac{1}{n} \sum_{j=1}^n \log\left(\frac{n}{2m} (F(x_{((j+m) \wedge n)} \nvert \hat{\mu}_n,\hat{\sigma}_n) - F(x_{((j-m) \vee 1)} \nvert \hat{\mu}_n,\hat{\sigma}_n))\right),
            \end{equation}
            where the window size $m$ is a positive integer such that $m < n/2$. The author is rather vague about the value of $m$, mentioning that the optimal choice of $m$ increases with $n$. Based on preliminary simulations, we recommend choosing $m=1$ if $n\leq 3$, $m=2$ if $n\in\{4, 5\}$ and in general $m=\text{round}((n+2)/5)$ for $n\geq 6$.\\

        \noindent \textbf{Alizadeh Noughabi-Park tests} These tests are based on different entropy estimators denoted $H_v$, $H_e$, $H_y$, $H_a$ and $H_z$. Let the window size $m$ be a positive integer such that $m < n/2$ (see Table~\ref{table.m.AP} for the specific choice of $m$ as a function of $n$) and define
                \begin{equation*}
                \hat{\theta}_{\beta}
                \leqdef
                \left\{\hspace{-1mm}
                \begin{array}{ll}
                    -\frac{1}{n} \sum_{i=1}^{n/2} \frac{\beta_i + \beta_{i+1}}{2} + \frac{1}{n} \sum_{i=n/2+1}^n \frac{\beta_i + \beta_{i+1}}{2}, &\mbox{if $n$ is even}, \\[3mm]
                    -\frac{1}{n} \sum_{i=1}^{(n-1)/2} \frac{\beta_i + \beta_{i+1}}{2} + \frac{\beta_{(n+1)/2+1} - \beta_{(n+1)/2}}{4n} + \frac{1}{n} \sum_{i=(n+1)/2+1}^n \frac{\beta_i + \beta_{i+1}}{2}, &\mbox{if $n$ is odd},
                \end{array}
                \right.
            \end{equation*}
            where the $\beta_i$'s are defined below for each entropy and are denoted by $\xi_i$, $\eta_i$, $\zeta_i$, $\nu_i$ or $\tau_i$.

            \vspace{3mm}
            A first entropy estimator, proposed in \cite{Vasicek_1976}, is
            \begin{equation}\label{eq:entropy.estimator.H.v}
                H_v \leqdef \frac{1}{n} \sum_{i=1}^n \log\left(\frac{n}{2m} (x_{((i + m) \wedge n)} - x_{((i - m) \vee 1)})\right),
            \end{equation}
            and the test statistic is
               \begin{equation*}
                    \text{AP}_v \leqdef \log(2 \hat{\theta}_{\xi}) + 1 - H_v,
            \end{equation*}
            where
            \begin{equation}\label{eq:definition.xi}
                \xi_i \leqdef \frac{1}{2m} \sum_{k=i-m}^{i+m-1} x_{((k\wedge n)\vee 1)}.
            \end{equation}

            \vspace{3mm}
            A second entropy estimator, proposed in \cite{Ebrahimi_Pflughoeft_Soofi_1994}, is
            \begin{equation}\label{eq:entropy.estimator.H.e}
                H_e \leqdef \frac{1}{n} \sum_{i=1}^n \log\left(\frac{n}{c_i m} (x_{((i+m) \wedge n)} - x_{((i-m) \vee 1)})\right),
            \end{equation}
            where
            \begin{equation*}
                c_i = 1 + \frac{\min\{i - 1, m, n - i\}}{m}.
            \end{equation*}
             Using this entropy estimator, the test statistic is
               \begin{equation*}
                    \text{AP}_e \leqdef \log(2 \hat{\theta}_{\eta}) + 1 - H_e,
            \end{equation*}
            where
            \begin{equation*}
                \eta_i \leqdef
                \left\{\hspace{-1mm}
                \begin{array}{ll}
                    \xi_{m+1} - \sum_{k=i}^m \frac{x_{(m+k)} - x_{(1)}}{m + k - 1}, &\mbox{if } 1 \leq i \leq m, \\[1mm]
                    \xi_i, &\mbox{if } m + 1 \leq i \leq n - m + 1, \\[1mm]
                    \xi_{n-m+1} + \sum_{k=n-m+2}^i \frac{x_{(n)} - x_{(k - m - 1)}}{n + m - k + 1}, &\mbox{if } n - m + 2 \leq i \leq n + 1,
                \end{array}
                \right.
            \end{equation*}
            and the $\xi_i$'s are defined in \eqref{eq:definition.xi}.

            \vspace{3mm}
            A third entropy estimator, proposed in \cite{Yousefzadeh_Arghami_2008}, is
            \begin{equation}\label{eq:entropy.estimator.H.y}
                    H_y
                    \leqdef \sum_{i=1}^n \left(\frac{\hat{F}_y(x_{((i+m) \wedge n)}) - \hat{F}_y(x_{((i-m) \vee 1)})}{\sum_{j=1}^n \hat{F}_y(x_{((j+m) \wedge n)}) - \hat{F}_y(x_{((j-m) \vee 1)})}\right)  \log\left(\frac{x_{((i+m) \wedge n)} - x_{((i-m) \vee 1)}}{\hat{F}_y(x_{((i+m) \wedge n)}) - \hat{F}_y(x_{((i-m) \vee 1)})}\right),
            \end{equation}
            where $\hat{F}_y$, an estimator of the cdf, can we written as
            \begin{equation*}
                \hat{F}_y(x_{(i)}) =
                \left\{\hspace{-2mm}
                \begin{array}{ll}
                    \frac{n-1}{n(n+1)} \hspace{-1mm}\left(\frac{n}{n-1} +
                    \left\{\hspace{-1.2mm}
                    \begin{array}{ll}
                        \frac{n}{2n-1}, &\mbox{if } x_{(1)} \neq x_{(2)} \\
                        0, &\mbox{if } x_{(1)} = x_{(2)}
                    \end{array}
                    \hspace{-1.5mm}\right\}\right)\hspace{-1mm}, &\mbox{if } i=1, \\[4mm]
                    \frac{n-1}{n(n+1)} \hspace{-1mm}\left(\frac{i(n-1)+1}{n-1} +
                    \left\{\hspace{-1.2mm}
                    \begin{array}{ll}
                        \frac{x_{(i)}-x_{(i-1)}}{x_{(i+1)}-x_{(i-1)}}, &\mbox{if } x_{(i-1)} \neq x_{(i+1)} \\
                        0, &\mbox{if } x_{(i-1)} = x_{(i+1)}
                    \end{array}
                    \hspace{-1.5mm}\right\}\right)\hspace{-1mm}, &\mbox{if }  2 \leq i \leq n-1, \\[4.7mm]
                    \frac{n-1}{n(n+1)} \hspace{-1mm}\left(\frac{(n-1)^2 + 1}{n-1} +
                    \left\{\hspace{-1.2mm}
                    \begin{array}{ll}
                        1 + \frac{n-1}{2n-1}, &\mbox{if } x_{(n-2)} \neq x_{(n-1)} \neq x_{(n)} \\
                        1, &\mbox{if } x_{(n-2)} \neq x_{(n-1)} = x_{(n)} \\
                        0, &\mbox{if } x_{(n-2)} = x_{(n)}
                    \end{array}
                    \hspace{-1.5mm}\right\}\right)\hspace{-1mm}, &\mbox{if } i=n.
                \end{array}
                \right.
            \end{equation*}
                Two tests based on this entropy have test statistics
                 \begin{equation*}
                    \text{AP}_y^{\scriptscriptstyle (\text{MLE})} \leqdef \log(2 \hat{\sigma}_n) + 1 - H_y
                \end{equation*}
            and
               \begin{equation*}
                    \text{AP}_y \leqdef \log(2 \hat{\theta}_{\zeta}) + 1 - H_y,
            \end{equation*}
            where
             \begin{equation*}
                \zeta_i \leqdef \frac{2m \xi_i}{\sum_{j=1}^n (\hat{F}_y(x_{((j+m) \wedge n)}) - \hat{F}_y(x_{((j-m) \vee 1)}))}, \quad 1 \leq i \leq n+1
            \end{equation*}
            and the $\xi_i$'s are defined in \eqref{eq:definition.xi}.

            \vspace{3mm}
            A fourth entropy estimator, proposed in \cite{Alizadeh_Noughabi_Arghami_2010}, is
            \begin{equation}\label{eq:entropy.estimator.H.a}
                H_a \leqdef \frac{1}{n} \sum_{i=1}^n \log\left(\frac{n}{a_i m} (x_{((i+m) \wedge n)} - x_{((i-m) \vee 1)})\right),
            \end{equation}
            where
            \begin{equation*}
                a_i \leqdef
                \left\{\hspace{-1mm}
\begin{array}{ll}
                    1, &\mbox{if } 1 \leq i \leq m \mbox{ or }n - m + 1 \leq i \leq n, \\[2mm]
                    2, &\mbox{if } m + 1 \leq i \leq n - m.
                \end{array}
                \right.
            \end{equation*}
            The test statistic based on this entropy estimator is
               \begin{equation*}
                    \text{AP}_a \leqdef \log(2 \hat{\theta}_{\nu}) + 1 - H_a,
            \end{equation*}
            where
            \begin{equation*}
                \nu_i \leqdef
                \left\{\hspace{-1mm}
                \begin{array}{ll}
                    \xi_{m+1} - \frac{1}{m} \sum_{k=i}^m (x_{(m+k)} - x_{(1)}), &\hspace{-2mm}\mbox{if } 1 \leq i \leq m, \\[1mm]
                    \xi_i, &\hspace{-2mm}\mbox{if } m + 1 \leq i \leq n - m + 1, \\[0mm]
                    \xi_{n-m+1} + \frac{1}{m} \sum_{k=n-m+2}^i (x_{(n)} - x_{(k - m - 1)}), &\hspace{-2mm}\mbox{if } n - m + 2 \leq i \leq n + 1
                \end{array}
                \right.
            \end{equation*}
            and the $\xi_i$'s are defined in \eqref{eq:definition.xi}.

            \vspace{3mm}
            Finally, a fifth entropy estimator, proposed in \cite{Zamanzade_Arghami_2011}, is
            \begin{equation}\label{eq:entropy.estimator.H.z}
                H_z \leqdef \frac{1}{n} \sum_{i=1}^n \log\left(\frac{n}{z_i m} (x_{((i+m) \wedge n)} - x_{((i-m) \vee 1)})\right),
            \end{equation}
            where
            \begin{equation*}
                z_i \leqdef
                \left\{\hspace{-1mm}
                \begin{array}{ll}
                    \frac{i}{m}, &\mbox{if } 1 \leq i \leq m, \\[1mm]
                    2, &\mbox{if } m + 1 \leq i \leq n - m, \\[1mm]
                    \frac{n - i + 1}{m}, &\mbox{if } n - m + 1 \leq i \leq n.
                \end{array}
                \right.
            \end{equation*}
            The test statistic based on this entropy estimator is
               \begin{equation*}
                    \text{AP}_z \leqdef \log(2 \hat{\theta}_{\tau}) + 1 - H_z,
            \end{equation*}
            where
            \begin{equation*}
                \tau_i \leqdef
                \left\{\hspace{-1mm}
                \begin{array}{ll}
                    \xi_{m+1} - \sum_{k=i}^m \frac{x_{(m+k)} - x_{(1)}}{k}, &\mbox{if } 1 \leq i \leq m, \\[1mm]
                    \xi_i, &\mbox{if } m + 1 \leq i \leq n - m + 1, \\[0mm]
                    \xi_{n-m+1} + \sum_{k=n-m+2}^i \frac{x_{(n)} - x_{(k - m - 1)}}{n - k + 2}, &\mbox{if } n - m + 2 \leq i \leq n + 1
                \end{array}
                \right.
            \end{equation*}
            and the $\xi_i$'s are defined in \eqref{eq:definition.xi}.

            \vspace{3mm}
            \noindent
            {\it Note 1:}
            Values of $m$ for $1\le n\le 120$, as proposed in \cite{Alizadeh_Noughabi_Park_2016}, are given in Table~\ref{table.m.AP}. We complete this table for $n>120$ by proposing extrapolations that satisfy $m/n \rightarrow 0$ as $n\rightarrow\infty$.
            \begin{table}[H]\centering
                \caption{Choice of $m$ as a function of sample size $n$ for the AP tests in (\ref{eq:entropy.estimator.H.v})--(\ref{eq:entropy.estimator.H.z})}
                \begin{tabular}{ccc}
                    \hline
                    & & \\[-3.4mm]
                    $n$ & $\text{AP}_v$, $\text{AP}_e$, $\text{AP}_y$, $\text{AP}_a$, $\text{AP}_y^{\scriptscriptstyle (\text{MLE})}$ & $\text{AP}_z$ \\
                    \hline
                    $~2 \leq n \leq 8~~$ & $1$ & $1$ \\
                    $~9 \leq n \leq 15~$ & $2$ & $2$ \\
                    $16 \leq n \leq 25~$ & $4$ & $2$ \\
                    $26 \leq n \leq 40~$ & $5$ & $3$ \\
                    $41 \leq n \leq 60~$ & $6$ & $4$ \\
                    $61 \leq n \leq 90~$ & $7$ & $5$ \\
                    $91 \leq n \leq 120$ & $8$ & $6$ \\
                    \hline
                    & & \\[-3.4mm]
                    $121 \leq n < \infty$ & $\text{round}((\log n)^{1.35})$ & $\text{round}((\log n)^{1.15})$ \\
                    \hline
                \end{tabular}\label{table.m.AP}
            \end{table}

            \vspace{1mm}
            \noindent
            {\it Note 2:} The authors also considered tests based on the statistic $\text{AP}_\ell^{\scriptscriptstyle (\text{MLE})} \leqdef \log(2 \hat{\sigma}_n) + 1 - H_\ell$ for $\ell\in\{v,e,a,z\}$. Whichever entropy estimator is used, these four tests generate exactly the same powers since they have the same variances. Furthermore, they are equivalent to the Choi-Kim $\text{CK}_v$ test in~\eqref{eq.Choi.tests}. Note that the $\text{AP}_v^{\scriptscriptstyle (\text{MLE})}$ based test was also proposed by \cite{Alizadeh_Noughabi_Arghami_2012}.\\

        \noindent \textbf{Choi-Kim tests} These three tests are based on entropy estimators in~\cite{Vasicek_1976}, \cite{Correa_1995} and \cite{van_Es_1992}. Their test statistics are
            \begin{equation}\label{eq.Choi.tests}
                \text{CK}_v \leqdef \frac{\exp\left(H_v\right)}{\hat{\sigma}_n},\quad \text{CK}_c \leqdef \frac{\exp\left(H_c\right)}{\hat{\sigma}_n},\quad
                \text{CK}_e \leqdef \frac{\exp\left(H_e\right)}{\hat{\sigma}_n},
            \end{equation}
            where
            \begin{align*}
                &H_v \leqdef \frac{1}{n} \sum_{i=1}^n \log\left(\frac{n}{2m} (x_{((i+m) \wedge n)} - x_{((i-m) \vee 1)})\right), \\
                &H_c \leqdef \frac{-1}{n} \sum_{i=1}^n \log\left(\frac{\sum_{j=i-m}^{i+m} (j/n - i/n) \big[x_{((j \wedge n) \vee 1)} - \frac{1}{2m+1} \sum_{k=i-m}^{i+m} x_{((k \wedge n) \vee 1)}\big]}{\sum_{j=i-m}^{i+m} \big[x_{((j \wedge n) \vee 1)} - \frac{1}{2m+1} \sum_{k=i-m}^{i+m} x_{((k \wedge n) \vee 1)}\big]^2}\right), \\
                &H_e \leqdef \frac{1}{n - m} \sum_{i=1}^{n - m} \log\Big(\frac{n+1}{m} (x_{((i+m) \wedge n)} - x_{(i)})\Big) + \sum_{k=m}^n \frac{1}{k} - \log\Big(\frac{n+1}{m}\Big),
            \end{align*}
            and the window size $m$ is a positive integer such that $m < n/2$. Values of $m$ for $1\le n\le 50$, as proposed in the study by \cite{Choi_Kim_2006}, are given in Table~\ref{table.m.CK}. In our power comparison, we extrapolated this table by choosing $m=10$ if $n=100$ and $m=20$ if $n=200$ for the $\text{CK}_v$ and $\text{CK}_c$ tests, and by choosing $m=2$ if $n\in\{100,200\}$ for the $\text{CK}_e$ test.

            \begin{table}[H]\centering
                \caption{Choice of $m$ as a function of sample size $n$ for the CK tests in \eqref{eq.Choi.tests}}
                \begin{tabular}{cccc}
                    \hline
                    $n$ & $\text{CK}_v$ & $\text{CK}_c$ & $\text{CK}_e$ \\
                    \hline
                    $~2 \leq n \leq 4~$ & $1$ & $1$ & $1$ \\
                    $~5 \leq n \leq 6~$ & $2$ & $2$ & $2$ \\
                    $~7 \leq n \leq 8~$ & $3$ & $3$ & $3$ \\
                    $~9 \leq n \leq 10$ & $3$ & $4$ & $4$ \\
                    $11 \leq n \leq 11$ & $3$ & $3$ & $5$ \\
                    $12 \leq n \leq 12$ & $3$ & $2$ & $2$ \\
                    $13 \leq n \leq 23$ & $3$ & $3$ & $2$ \\
                    $24 \leq n \leq 25$ & $4$ & $3$ & $2$ \\
                    $26 \leq n \leq 33$ & $4$ & $4$ & $2$ \\
                    $34 \leq n \leq 37$ & $5$ & $4$ & $2$ \\
                    $38 \leq n \leq 46$ & $5$ & $5$ & $2$ \\
                    $47 \leq n \leq 50$ & $6$ & $5$ & $2$ \\
                    \hline
                \end{tabular}\label{table.m.CK}
            \end{table}

    \subsection{Other tests}\label{sec:tests.other}

        \noindent \textbf{Alizadeh Noughabi-Balakrishnan tests} These five tests are based on the original test statistics of \cite{Csiszar_1963}, \cite{Morimoto_1963} and \cite{Ali_Silvey_1966}, where the unknown location and scale parameters, $\mu$ and $\sigma$, are replaced by their ML estimators. More precisely, they are based on an estimator $\hat{D}_{\nu}$ of Csiszar's divergence given by
            \begin{equation*}
                D_{\nu}(g, f) \leqdef \int_{-\infty}^{\infty}f(x\nvert \mu, \sigma)\,\nu\left(\frac{g(x)}{f(x\nvert \mu, \sigma)}\right)dx,
            \end{equation*}
            where $g(x)$ is the true density of the observations $x_1,\dots,x_n$, $f(x\nvert \mu, \sigma)$ is the Laplace density and $\nu(\cdot)$ is a divergence measure. The estimator of $D_{\nu}(g, f)$ is defined as
            \begin{equation*}
                \hat{D}_{\nu} = \frac{1}{n} \sum_{i=1}^n \frac{f(x_i \nvert \hat{\mu}_n, \hat{\sigma}_n)}{\hat{g}(x_i)} \nu\left(\frac{\hat{g}(x_i)}{f(x_i \nvert \hat{\mu}_n, \hat{\sigma}_n)}\right),
            \end{equation*}
            where
            \begin{equation*}
                \hat{g}(x_i) \leqdef \frac{1}{n \hspace{0.3mm}\hat{h}} \sum_{j=1}^n k\left(\frac{x_i - x_j}{\hat{h}}\right), \quad k(z) \leqdef \phi(z), \quad \hat{h} \leqdef \hat{\sigma}_n \left(\frac{2}{n \sqrt{\pi}}\right)^{1/5}.
            \end{equation*}
            The test statistics are
            \begin{equation}\label{eq:sec:tests.other.Alizadeh.Noughabi.Balakrishnan}
            \begin{aligned}
                \text{AB}_{\text{KL}} &\leqdef \hat{D}_{\nu}, \quad \text{with } &&\nu(t) = t \log t \text{ (Kullback-Leibler divergence)}, \\[-0.2mm]
                \text{AB}_{\text{He}} &\leqdef \hat{D}_{\nu}, \quad \text{with } &&\nu(t) = \tfrac{1}{2} (\sqrt{t} - 1)^2 \text{ (Hellinger distance)}, \\[-0.2mm]
                \text{AB}_{\text{Je}} &\leqdef \hat{D}_{\nu}, \quad \text{with } &&\nu(t) = (t - 1) \log t \text{ (Jeffreys distance)}, \\[-0.2mm]
                \text{AB}_{\text{TV}} &\leqdef \hat{D}_{\nu}, \quad \text{with } &&\nu(t) = |t - 1| \text{ (Total variation distance)}, \\[-0.2mm]
                \text{AB}_{\text{$\chi^2$}} &\leqdef \hat{D}_{\nu}, \quad \text{with } &&\nu(t) = (t - 1)^2\,(\chi^2\text{-divergence)}.
              \end{aligned}
              \end{equation}

            \vspace{3mm}
            \noindent {\it Note:} We chose the bandwidth $h$ that is optimal under $H_0$ (see Equation (3.21) in \cite{Silverman_1986}), namely
            \begin{equation*}
                h = \left\{\frac{\int_{-\infty}^{\infty} (k(z))^2 \rd z}{n \int_{-\infty}^{\infty} \left(\frac{\rd^2}{\rd x^2} f(x \nvert \mu, \sigma)\right)^2 \rd x}\right\}^{1/5} \hspace{-2mm}=\hspace{1mm} \left\{\frac{\frac{1}{2\sqrt{\pi}}}{n \cdot \frac{1}{4 \sigma^5}}\right\}^{1/5} \hspace{-2mm} = \hspace{1mm} \sigma \left(\frac{2}{n \sqrt{\pi}}\right)^{1/5}.
            \end{equation*}
            If the chosen kernel function were $k(z) \leqdef \psi(z)$ instead, we would obtain $h= \sigma n^{-1/5}$.
            In \cite{Alizadeh_Noughabi_Balakrishnan_2016}, the chosen bandwidth is $h=1.06 \sigma n^{-1/5}$ and it is estimated by $\hat{h}=1.06 s_n n^{-1/5}$, but this bandwidth is only optimal when the distribution under $H_0$ is normal.\\

        \noindent \textbf{Alizadeh Noughabi-Jarrahiferiz test} This test is based on the informational energy estimator of \cite{Pardo_2003} based on spacing of order statistics for the sequence of observations $\hat{u}_1, \hat{u}_2, \dots, \hat{u}_n$. Its test statistic is
            \begin{equation}\label{eq:original:sec:tests.other.Alizadeh.Noughabi.Jarrahiferiz}
                \text{AJ} = \frac{1}{n} \sum_{j=1}^n \frac{2m}{n(F(x_{((j+m) \wedge n)} \nvert \hat{\mu}_n,\hat{\sigma}_n) - F(x_{((j-m) \vee 1)} \nvert \hat{\mu}_n,\hat{\sigma}_n))},
            \end{equation}
            where the window size $m$ is a positive integer such that $m < n/2$. Values of $m$ for $1\le n\le 120$,  as proposed in \cite{Alizadeh_Noughabi_Jarrahiferiz_2019}, are given in Table~\ref{table.m.AJ}. We complete this table for $n>120$ by proposing an extrapolation which satisfies $m/n \rightarrow 0$ as $n\rightarrow\infty$.

            \begin{table}[H]\centering
                \caption{Choice of $m$ as a function of sample size $n$ for the AJ test in \eqref{eq:original:sec:tests.other.Alizadeh.Noughabi.Jarrahiferiz}}
                \begin{tabular}{cc}
                    \hline
                    $n$ & $m$ \\
                    \hline
                    $~1 \leq n \leq 8~~$ & $2$ \\
                    $~9 \leq n \leq 15~$ & $3$ \\
                    $16 \leq n \leq 25~$ & $5$ \\
                    $26 \leq n \leq 35~$ & $6$ \\
                    $36 \leq n \leq 45~$ & $7$ \\
                    $46 \leq n \leq 60~$ & $8$ \\
                    $61 \leq n \leq 90~$ & $9$ \\
                    $91 \leq n \leq 120$ & $10$ \\
                    \hline
                    & \\[-3.4mm]
                    $121 \leq n < \infty$ & $\text{round}((\log n)^{1.5})$ \\
                    \hline
                \end{tabular}\label{table.m.AJ}
            \end{table}

        \noindent \textbf{Brain-Shapiro test (adapted by Gulati)} This test is adapted from the regression test of \cite{Brain_Shapiro_1983}, where the case of the exponential distribution was treated under $H_0$. Its test statistic is
            \begin{equation}\label{eq:sec:tests.other.Brain.Shapiro}
                \text{BS} \leqdef 12 (n-1) (\overline{v} - 1/2)^2 + \frac{5 (n-1)}{(n+2)(n-2)} \left(n - 2 + 6 n \overline{v} - 12 \sum_{i=1}^{n-1} \frac{i v_i}{n - 1}\right)^2\hspace{-1mm},
            \end{equation}
            where
            \begin{equation*}
                \begin{aligned}
                    &\overline{v} \leqdef \frac{1}{n-1} \sum_{i=1}^{n-1} v_i, \quad v_i \leqdef \frac{\sum_{j=1}^i w_j}{\sum_{j=1}^{n} w_j},\\
                    &w_j \leqdef (n - j + 1)(y_{(j)} - y_{(j - 1)}), \quad y_j \leqdef |\hat{z}_j|, ~~\text{with } y_{(0)} \leqdef 0.
                \end{aligned}
            \end{equation*}

            \noindent{\it Note:}  We made a correction in \eqref{eq:sec:tests.other.Brain.Shapiro}, where we used the fraction $\frac{5 (n-1)}{(n+2)(n-2)}$ as in \cite{Brain_Shapiro_1983}  (equation (6) with $m$ set to $n+1$) instead of $\frac{5 (n-1)}{(n+1)(n-2)}$ as in \cite{Gulati_2011} (equation (2.3)).\\

        \noindent \textbf{Kozubowski-Panorska test} This test is based on the original test statistic of \cite{Kozubowski_Panorska_2004}, which is a two-tailed likelihood ratio test of symmetry  for the asymmetric Laplace distribution, with unknown location parameter $\mu$ estimated by $\hat{\mu}_n$. Its test statistic is
            \begin{equation}\label{eq:sec:tests.other.Kozubowski.Panorska}
                \text{KP} \leqdef n \, \left(2 - \frac{\big(1 + \sqrt{\hat{k}^4}\big)^2}{1 + \hat{k}^4}\right),
            \end{equation}
            where
            \begin{equation*}
                \hat{k}^4 \leqdef \frac{\frac{1}{n} \sum_{i=1}^n \max\{-(x_i - \hat{\mu}_n),0\}}{\frac{1}{n} \sum_{i=1}^n \max\{x_i - \hat{\mu}_n,0\}}.
            \end{equation*}

            \vspace{3mm}\noindent{\it Note 1:} A positive (negative) value of $\hat{k}^4 - 1$ is associated with a left-skewed (right-skewed) distribution.

            \vspace{3mm}\noindent{\it Note 2:} It can be shown that
            \begin{equation*}
                \frac{1}{n} \sum_{i=1}^n \max\{-(x_i - \hat{\mu}_n),0\} - \frac{1}{n} \sum_{i=1}^n \max\{x_i - \hat{\mu}_n,0\}=\hat{\mu}_n-\bar{x}_n.
            \end{equation*}
            Given that the asymmetry component of the $\text{DLO}_{X}$ test given by $Z_1$ (see Equation~\eqref{eq:tests.moments.Desgagne.Lafaye.Ouimet}) is a z-score also based on $\hat{\mu}_n-\bar{x}_n$, the two test statistics $Z_1$ and KP lead to virtually identical $p$-values for all samples.\\

        \noindent \textbf{Meintanis tests} These two tests are based on a weighted measure (involving a parameter $a$) of relative deviation between the empirical characteristic function ($n^{-1} \sum_{j=1}^n e^{\ii t \hat{z}_j}$) and the characteristic function under $H_0$, for the sample standardized by ML estimators. Standardization of the observations based on the MOM is also proposed but not recommended by the author, because his simulation study suggests that the ML estimators are more robust with respect to $a$. Consider
            \begin{equation}\label{eq:original:sec:tests.empirical.characteristic.function.Meintanis}
                \text{Me}_a^{(j)} = n \int_{-\infty}^{\infty} D_n^2(t) w_j(t,a) dt,
            \end{equation}
            where $D_n(t) \leqdef |(1 + t^2)\frac{1}{n} \sum_{j=1}^n e^{\ii t \hat{z}_j} - 1|$ and the two proposed weight functions are $w_1(t,a) = \exp(-a |t|)$ and $w_2(t,a) = \exp(-at^2)$, for $a > 0$. The test statistics are
            \begin{align*}
                \text{Me}_a^{(1)}
                &\leqdef \frac{2n}{a} - 4 a \sum_{j=1}^n
                    \left[\hspace{-1mm}
                        \begin{array}{l}
                            \frac{1}{a^2 + \hat{z}_j^2} + \frac{2(a^2 - 3 \hat{z}_j^2)}{(a^2 + \hat{z}_j^2)^3}
                        \end{array}
                        \hspace{-1mm}\right] + \frac{2a}{n} \sum_{j,k=1}^n
                    \left[\hspace{-1mm}
                    \begin{array}{l}
                        \frac{1}{a^2 + (\hat{z}_j - \hat{z}_k)^2} + \frac{4(a^2 - 3(\hat{z}_j - \hat{z}_k)^2)}{(a^2 + (\hat{z}_j - \hat{z}_k)^2)^3} \\[1.5mm]
                        + \frac{24(a^4 + 5(\hat{z}_j - \hat{z}_k)^4 - 10 a^2 (\hat{z}_j - \hat{z}_k)^2)}{(a^2 + (\hat{z}_j - \hat{z}_k)^2)^5}
                    \end{array}
                    \hspace{-1mm}\right],  \\
                \text{Me}_a^{(2)}
                &\leqdef n\sqrt{\frac{\pi}{a}} - 2\sqrt{\frac{\pi}{a}} \sum_{j=1}^n
                    \left[\hspace{-1mm}
                        \begin{array}{l}
                            1 - \frac{\hat{z}_j^2 - 2a}{4 a^2}
                        \end{array}
                        \hspace{-1mm}\right] e^{-\hat{z}_j^2/4a} \notag \\
                &\quad+ \frac{2}{n} \sqrt{\frac{\pi}{a}} \sum_{j,k=1}^n
                    \left[\hspace{-1mm}
                    \begin{array}{l}
                        \frac{1}{2} - \frac{(\hat{z}_j - \hat{z}_k)^2 - 2a}{4 a^2} \\[1.5mm]
                        + \frac{((\hat{z}_j - \hat{z}_k)^4 + 12 a^2 - 12 a (\hat{z}_j - \hat{z}_k)^2)}{32 a^4}
                    \end{array}
                    \hspace{-1mm}\right] e^{-(\hat{z}_j - \hat{z}_k\hspace{-0.2mm})^2/4a}.
            \end{align*}

            \noindent
            {\it Note:}
            The parameter $a$ controls the rate of decay of the weight function. In his simulation study, \cite{Meintanis_2004} considered $a\in \{0.5,1,2,4,5,10\}$ and finally recommended $\text{Me}_2^{\scriptscriptstyle (1)}$ and $\text{Me}_{0.5}^{\scriptscriptstyle (2)}$ as the best choices. In a preliminary power study, we came to the same conclusion. Therefore, we chose $\text{Me}_2^{\scriptscriptstyle (1)}$ and $\text{Me}_{0.5}^{\scriptscriptstyle (2)}$  for our power comparison.\\

        \noindent \textbf{Subramanian-Dixit test} This is a two-tailed likelihood ratio test of symmetry for the asymmetric Laplace distribution. The test statistic is
            \begin{equation}\label{eq:sec:tests.other.Subramanian}
                \text{SD} \leqdef \frac{U}{U+V},
            \end{equation}
            where
            \begin{equation*}
                U \leqdef \sum_{i=1}^{n_1} (x_{(n_1)} - x_{(i)}) \quad\text{and}\quad V \leqdef \sum_{i=n_1+1}^n (x_{(i)} - x_{(n_1+1)}),
            \end{equation*}
            \begin{equation*}
                n_1 \leqdef \max\{i\in \{1,2,\dots,n\} : x_i \leq \hat{\theta}_n\}\in\{1,2,\dots,n\},
            \end{equation*}
            with $U\leqdef 0$ if $n_1=1$, $V \leqdef 0$ if $n_1=n$ and $\text{SD} \leqdef 0.5$ if $U=V=0$, where $\hat{\theta}_n$ is the half-sample mode estimator from \cite{Robertson_Cryer_1974}, implemented in the function \texttt{hsm()} of the \texttt{R} package \texttt{modeest}. Note that any other nonparametric estimator of the mode could be used instead of $\hat{\theta}_n$, as suggested in \citet[p.5]{Subramanian_Dixit_2018}.

            \vspace{3mm}\noindent
            {\it Note:} The range of the test statistic is $0\le\text{SD}\le 1$, with a small (large) value associated with a right-skewed (left-skewed) distribution.\\

        \noindent \textbf{Sz\'ekely-Rizzo test (applied by Rizzo and Haman)} This test, based on energy distance (see e.g.\ \cite{Szekely_Rizzo_2013}), is an adaptation of the original statistic in \cite{Szekely_Rizzo_2005} defined by
            \begin{equation}\label{eq:original:sec:tests.other.Szekely.Rizzo}
                \text{SR} \leqdef 2 \sum_{i=1}^n \EE|x_i - X| - n \EE|X - X'| - \frac{2}{n} \sum_{k=1}^n (2k-1-n)x_{(k)},
            \end{equation}
            where $X,X'\sim G$ i.i.d. for a given distribution $G$.

            In \cite{Rizzo_Haman_2016}, the expectations were calculated explicitly for the asymmetric Laplace distribution. In particular, when $Z,Z'\sim \text{Laplace}(0,1)$ are i.i.d., we have $\EE|z - Z| = |z| + e^{-|z|}$ for any point $z\in \R$, and $\EE|Z - Z'| = 3/2$. As the test considered by the authors is $H_0: X_i\sim \text{Laplace}(\mu = 0,\sigma = 1)$, we normalize the observations from $x_i$ to $\hat{z}_i$ to allow for the fact that here $\mu$ and $\sigma$ are assumed to be unknown. The test statistic is then given by
            \begin{equation*}
                \text{SR}^{\star} \leqdef 2 \sum_{i=1}^n \left(|\hat{z}_i| + e^{-|\hat{z}_i|}\right) - \frac{3n}{2} - \frac{2}{n} \sum_{k=1}^n (2k-1-n)\hat{z}_{(k)}.
            \end{equation*}

\section{Empirical power comparison}\label{sec:empirical.power.comparison}

    \subsection{Design of the Monte Carlo study}\label{sec:Design.of.the.MonteCarlo.study}

        In order to fully explore the power characteristics of goodness-of-fit tests, empirical power comparisons usually aim to include a wide range of statistical distributions as alternatives to that under the null hypothesis. For example, in the context of goodness-of-fit tests for the normal distribution, \cite{Romao_Delgado_Costa_2010} selected 79 particular cases of 16 models (beta, Cauchy, Laplace, logistic, Student-$t$, Tukey, $\chi^2$, gamma, Gumbel, log-normal, Weibull, truncated normal, location contaminated normal, scale contaminated normal, mixture of normals, normal with outliers) as alternative distributions while \cite{Desgagne_Lafaye_2018} considered 85 particular cases of 10 models (asymmetric exponential power, symmetrical exponential power, Student-$t$, logistic, beta, $\chi^2$, gamma, Gumbel, log-normal and Weibull) as alternative distributions.

        In our simulation study we included most of the alternative models used in the literature for the empirical power comparison of Laplace tests. Specifically, we used the 11 models listed in Table~\ref{table:distribution}. As the 40 test statistics are location and scale invariant, all distributions referred to there are standardized versions of the various distributions, with location and scale parameters set to 0 and 1, respectively. Several classical distributions which, at first sight, might be considered not to have been included in our study are in fact special cases of these 11 models: the Laplace distribution as $\text{Laplace}(0,1)\equiv\text{GED}(1)\equiv\text{MixL}(\cdot,0,1)\equiv\text{ALp}(1)$, the normal distribution as $\textrm{N}(0,1)\equiv\sqrt{2}\,\text{GED}(2)\equiv\text{MixN}(\cdot,0,1)\equiv\text{SkewN}(0)$, the Cauchy distribution as $\text{Cauchy}(0,1)\equiv t(1)$, the logistic distribution as $\text{Logistic}(0,1)\equiv\text{Tu}(0)$, the exponential distribution as $\text{Exp}(1)\equiv \textrm{G}(1)\equiv \textrm{W}(1)$, the chi-squared distribution as $\chi^2(k)\equiv 2\, \textrm{G}(k/2)$ and the uniform distribution as $\textrm{U}(-1,1)\equiv\text{Tu}(1)$. The inverse Gaussian distribution was not included in our study but it generated very similar power curves to those for the gamma distribution, probably because both are very similar in terms of excess skewness and kurtosis. Although many other classical statistical models exist, the careful identification of 400 particular cases of the 11 models led to a wide-ranging set of alternative distributions.

        \def\arraystretch{1.1}
        \begin{table}[H]\centering
        \scriptsize
        \caption{The 11 (standardized) alternative models used in the simulation study.}
\\[3mm]
                \hline
                && \\[-3mm]
                Weibull
                & $\textrm{W}(k)$, $k > 0$
                & $k x^{k-1} \exp(-x^k)$ for $x > 0$\\[1mm]
                \hline
            \end{tabular}
            \label{table:distribution}
        \end{table}

        To build such a set, we identified 20 submodels drawn from the 11 main models, consisting of 12 symmetric alternatives -- divided equally into 6 heavy-tailed and 6 light-tailed distributions (see Table~\ref{table:symmetric.alternatives}) -- and 8 asymmetric alternatives (see Table~\ref{table:asymmetric.alternatives}). Note that the classification into heavy-tailed versus light-tailed is relative to the Laplace distribution's tailedness. Consider, for example, the asymmetric Laplace distribution $\text{ALp}(k)$ in Table~\ref{table:asymmetric.alternatives}, where the parameter $k$ controls the degree of asymmetry. For each sample size, we carefully selected the smallest interval for $k$ under which the entire range of the power curve space is covered for a large pre-selection of tests that are known to perform well. We obtained $k\in(\mathbf{1}, 5]$ for $n=20$, $k\in(\mathbf{1}, 3]$ for $n=50$, $k\in(\mathbf{1}, 2.5]$ for $n=100$ and $k\in(\mathbf{1}, 2]$ for $n=200$. The value displayed in boldface corresponds to the alternative closest to the Laplace distribution, which means its power is the closest to the significance level. The other endpoint of the interval corresponds to the alternative furthest away from a Laplace distribution, and to the largest power. Then we evenly divided each parameter interval (excluding the value in bold) to obtain 20 specific distributions per submodel. For example, for $n=50$, the $\text{ALp}(1.1)$, $\text{ALp}(1.2),\dots,\text{ALp}(2.9)$, $\text{ALp}(3)$ distributions were used as the specific distributions from the asymmetric Laplace model. For this distribution, considering values of $k$ beyond 3 would not provide additional information because the power curves become flat for most tests. Using 20 specific distributions for each of the 20 submodels led to a total of 400 alternatives to the Laplace distribution. Note that the parameter ranges used were the same regardless of the significance level because their impact on the selection process was negligible.

        \def\arraystretch{1.5}
        \begin{table}[H]\centering
        \scriptsize
        \caption{The 12 symmetric heavy/light-tailed submodels.\\ The numbers in curly brackets correspond to the sample sizes $n=20,50,100$ and $200$, respectively.}
\begin{tabular}{lll}
                \hline
                \textbf{Distribution} & \textbf{Heavy tails} & \textbf{Light tails} \\
                \hline
                $\text{GED}(k)$ & $k\in[0.1,\mathbf{1})$ for all $n$ & $k\in(\mathbf{1},b]$, $b\in\{10, 6.2, 4.8, 3\}$ \\
                \hline
                $\text{MixL}(\omega=0.5,k_1,k_2)$   & $k_1=0, k_2\in(\mathbf{1}, b]$, $b\in\{50, 20, 12, 8\}$  & $k_1\in(\mathbf{0},b]$, $b\in\{12,6,4,3\}$, $k_2=1$ \\
                \hline
                $\text{MixN}(\omega=0.5,k_1,k_2)$  & $k_1=0, k_2\in(\mathbf{3.5}, b]$, $b\in\{40, 20, 12, 8\}$  & $k_1\in(\mathbf{0},b]$, $b\in\{10, 5, 3, 3\}$, $k_2=1$ \\
                \hline
                $\text{NIG}(k_1,k_2=0)$  & $k_1\in[0.01, \mathbf{b})$, $\mathbf{b}\in\{0.6, 0.7, 0.7, 0.65\}$  & $k_1\in(\mathbf{0.7},10]$ for all $n$ \\
                \hline
                $t(k)$ & $k\in[0.5, \mathbf{3.5})$ for all $n$  & $k\in(\mathbf{3.5},b]$, $b\in\{30,30,30,20\}$ \\
                \hline
                $\text{Tu}(k)$ & $k\in[-a, -\mathbf{b})$, $a\in\{1.5,1.5,1.5,1.2\}$ & $k\in(-\mathbf{a}, b]$, $\mathbf{a}\in\{0.14,0.14,0.18,0.16\}$  \\
                & \hspace{17.4mm}$\mathbf{b}\in\{0.14,0.14,0.18,0.20\}$ & \hspace{15.3mm}$b\in\{1.65,1.5,1.5,0.4\}$ \\
                \hline
            \end{tabular}
            \label{table:symmetric.alternatives}
        \end{table}

        \vspace{-2mm}
        \def\arraystretch{1.5}
        \begin{table}[H]\centering
        \scriptsize
        \caption{The 8 asymmetric submodels.\\The numbers in curly brackets correspond to the sample sizes $n=20,50,100$ and $200$, respectively.}
            \begin{tabular}{ll}
                \hline
                \textbf{Distribution} & \textbf{Range of the parameters} \\
                \hline
                $\text{ALp}(k)$ & $k\in(\mathbf{1}, b]$, $b\in\{5, 3, 2.5, 2\}$ \\
                \hline
                $\text{G}(k)$ & $k\in[0.5, \mathbf{b})$, $\mathbf{b}\in\{7, 10.5, 15, 15\}$ \\
                \hline
                $\text{LN}(k)$ & $k\in(\mathbf{0.1}, b]$, $b\in\{3, 2, 2, 2\}$ \\
                \hline
                $\text{MixL}(\omega=0.25,k_1,k_2=1)$  & $k_1\in(\mathbf{0},b]$, $b\in\{15, 8, 6, 4\}$ \\
                \hline
                $\text{MixN}(\omega=0.25,k_1,k_2=1)$  & $k_1\in(\mathbf{0},b]$, $b\in\{15, 10, 6, 4\}$ \\
                \hline
                $\text{NIG}(k_1=0.5,k_2)$  & $k_2\in(\mathbf{0}, 0.49]$ for all $n$ \\
                \hline
                $\text{SkewN}(k)$ & $k\in(\mathbf{0}, b]$, $b\in\{15, 15, 10, 6\}$ \\
                \hline
                $\text{W}(k)$ & $k\in[0.1, \mathbf{b})$, $\mathbf{b}\in\{3, 5, 8, 5\}$ \\
                \hline
           \end{tabular}
           \label{table:asymmetric.alternatives}
        \end{table}

        Our Monte Carlo simulation study was carried out for sample sizes of $n = 20, 50, 100$ and $200$, and for significance levels of $\alpha = 0.01, 0.05$ and $0.10$. For each specific scenario, the empirical power of each of the 40 Laplace tests was measured by the proportion of 100,000 simulated samples for which the Laplace composite hypothesis was rejected. We used critical values for each test computed from 1,000,000 samples simulated from the Laplace distribution, so as to ensure that the empirical levels were very close to the nominal significance levels. In total, we obtained $400\times 4\times 3 = 4,800$ empirical power estimates.

        All the simulations were carried out on a Linux computing cluster equipped with 10 cores and 124 GB of RAM. We used version~3.6.1 of the \RR software and version~1.1.2 of the \RR\!\!\texttt{/C++} \texttt{PoweR} package \citep{Tran_2013, Lafaye_de_Micheaux_Tran_2016}, freely available on the CRAN \citep{Rsoftware}. Overall, our simulations took approximately 32 hours. The R/C++ code used to obtain the numerical results is available at \url{http://biostatisticien.eu/JSCS2022}.

    \subsection{Analysis of the results from the Monte Carlo study}\label{sec:empirical.power.results}

        Table~\ref{table:critical.values.A.1}, in \ref{sec:power.tables.complete}, gives the critical values used for each of the 40 tests, for the sample sizes $n = 20, 50, 100, 200$, and the significance levels $\alpha = 0.01, 0.05, 0.10$.

        The raw power results from the simulations are provided in XLS format downloadable from \url{http://biostatisticien.eu/JSCS2022}.
        For each one of the 12 combinations of $\alpha$ and $n$, we averaged the powers of each test against each of the 20 submodels. For $\alpha=0.05$ and $n=20, 50, 100, 200$, these results are given in Tables~\ref{new:table:alpha.0.05.n.20}--\ref{new:table:alpha.0.05.n.200} respectively. For $\alpha=0.01, 0.10$, the analogous results are given in the supplementary materials document, in Tables~\ref{new:table:alpha.0.01.n.20}--\ref{new:table:alpha.0.01.n.200} and Tables~\ref{new:table:alpha.0.10.n.20}--\ref{new:table:alpha.0.10.n.200}, respectively.

        The power results in the 12 tables were averaged again for 5 different groups of alternative distributions: all 20 submodels, the 12 symmetric submodels, the 6 symmetric heavy-tailed submodels, the 6 symmetric light-tailed submodels, and the 8 asymmetric submodels. For $\alpha=0.05$ and $n=20, 50, 100, 200$, the results are presented in Tables~\ref{new:table:alpha.0.05.n.20.specialized.ranks}--\ref{new:table:alpha.0.05.n.200.specialized.ranks} respectively. For $\alpha=0.01, 0.10$, the analogous results are given  in the supplementary materials document, in Tables~\ref{new:table:alpha.0.01.n.20.specialized.ranks}--\ref{new:table:alpha.0.01.n.200.specialized.ranks} and in Tables~\ref{new:table:alpha.0.10.n.20.specialized.ranks}--\ref{new:table:alpha.0.10.n.200.specialized.ranks}, respectively.

        An analysis of the results for each of these 5 groups of alternatives is given in the following 5 subsections. The analysis corresponds to $\alpha=0.05$, but is also valid for $\alpha = 0.01, 0.10$.

   \subsubsection{The best omnibus tests}\label{sec:all.alternative}

        In this section, we identify the best omnibus tests against all of the alternative distributions considered in the simulation study, irrespective of whether they were symmetrical/asymmetrical light/heavy-tailed. To achieve this, in Table~\ref{table:overall.average.power.n20.to.200} we present results for the 10 best performing tests against the ``All'' grouping of alternative distributions in Tables~\ref{new:table:alpha.0.05.n.20.specialized.ranks}--\ref{new:table:alpha.0.05.n.200.specialized.ranks}. As an interpretational aid, we define the ``gap'' as the difference between the maximum average power amongst the 40 tests and the average power of a given test. We obtained four such gaps for each test, one for each sample size. The ``maximum gap'' and ``average gap'' for a given test are defined as the maximum and the average of those 4 gaps. To complete Table~\ref{table:overall.average.power.n20.to.200}, we have included the 10 best tests in terms of maximum and average gaps.

        \bgroup
        \def\arraystretch{1.0}
        \begin{table}[H]\centering
        \scriptsize
            \caption{The average \% power of the 10 best performing tests for the ``All'' group of alternatives extracted from Tables~\ref{new:table:alpha.0.05.n.20.specialized.ranks}--\ref{new:table:alpha.0.05.n.200.specialized.ranks}, as a function of sample size, $n$.}\label{table:overall.average.power.n20.to.200}
            \begin{center}
                \setlength\tabcolsep{4.87pt} 
\begin{tabular}{ccccccccccc}
                \hline
                $n = 20$ &  {\bf $\text{AP}_y$} &  {\bf $\text{AP}_e$} &  {\bf $\text{AP}_z$} & {\bf $\text{AP}_v$} & {\bf  $\text{CK}_v$} & {\bf $\text{AP}_y^{\normalfont \text{(MLE)}}$} & {\bf $\text{Me}_2^{(1)}$} & {\bf $\text{AP}_a$} & {\bf Wa} & {\bf $\text{DLO}_X$} \\
                {\bf Power} &  47.0 &  46.4 &  46.2 & 45.3 & 45.2 & 44.9 & 44.7 & 44.5 & 44.2 & 44.0 \\
                {\bf Gap} &  - &  0.6 &  0.8 & 1.7 & 1.8 & 2.1 & 2.3 & 2.5 & 2.8 & 3.0 \\
                \hline
                $n = 50$ &  {\bf $\text{AP}_v$} & {\bf $\text{AP}_e$} & {\bf $\text{AP}_y^{\scriptscriptstyle (\text{MLE})}$} & {\bf $\text{DLO}_X$} & {\bf $\text{AP}_y$} & {\bf $\text{Me}_{0.5}^{\scriptscriptstyle (2)}$} & {\bf $\text{Me}_2^{\scriptscriptstyle (1)}$} & {\bf LK} & {\bf  Wa} & {\bf $\text{CK}_v$} \\
                {\bf Power} &  63.7 & 62.6 & 62.3 & 60.8 & 60.4 & 60.4 & 60.1  & 59.0 & 58.9 & 58.8 \\
                {\bf Gap} &  - & 1.1 & 1.4 & 2.8 & 3.2 & 3.3 & 3.5 & 4.7 & 4.7 & 4.8 \\
                \hline
                $n = 100$ &  {\bf $\text{DLO}_X$} & {\bf LK} & {\bf Wa} & {\bf $\text{AP}_v$} & {\bf $\text{Me}_{0.5}^{\scriptscriptstyle (2)}$} & {\bf $\text{Me}_2^{\scriptscriptstyle (1)}$} & {\bf $\text{AP}_y^{\normalfont \text{(MLE)}}$} & {\bf $\text{AP}_e$} & {\bf $\text{AB}_{\text{He}}$} & {\bf Ku} \\
                {\bf Power} &  74.3 & 72.3 & 71.8 & 71.7 & 71.7 & 71.4 & 69.0 & 68.6 & 68.1 & 67.8 \\
\hline
                \end{tabular}
            \end{center}
        \end{table}
        \egroup

        The best omnibus tests by sample size are those based on: $\text{AP}_y$ for $n=20$, $\text{AP}_v$ for $n=50$, and $\text{DLO}_X$ for $n=100,200$. The best omnibus test regardless of sample size is the $\text{DLO}_{X}$ test, with an average gap of 1.5\% and a maximum gap of 3\%.
        We observe that all four groups of tests are represented among the 10 best tests, which suggests that no single approach is superior and that the quality of the design of a test is more crucial. Indeed, the $\text{DLO}_{X}$ and LK tests are based on moments, the $\text{Me}_2^{\scriptscriptstyle (1)}$ and $\text{Me}_{0.5}^{\scriptscriptstyle (2)}$  tests are based on the empirical characteristic function, the $\text{AP}_v$ test is based on entropy, and the Wa test is based on the empirical cumulative distribution function.

\begin{figure}[ht]\centering
            \begin{center}
                \begin{tabular}{cc}
                    \includegraphics[width=0.60\textwidth]{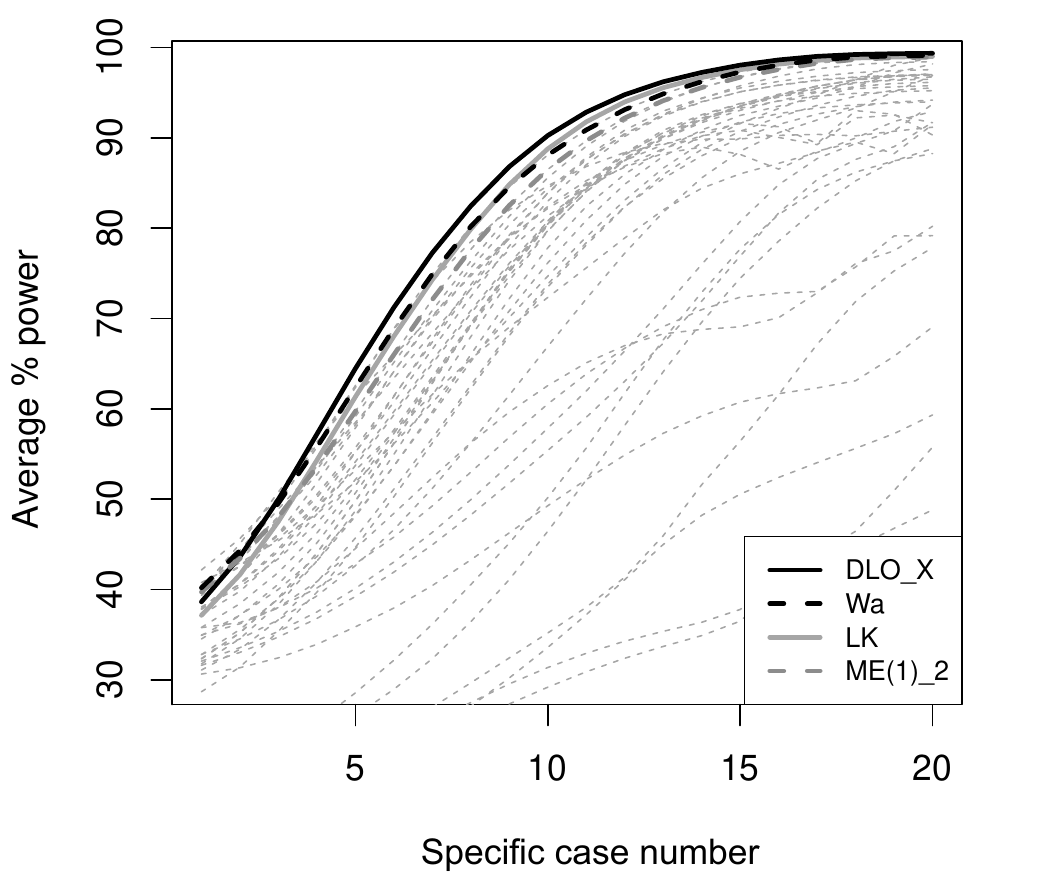}
                \end{tabular}
            \end{center}
            \vspace{-4mm}
            \caption{Power curves of the averaged \% power of the 40 tests for $\alpha=0.05$ and $n=200$. Only the four curves with the largest overall power are highlighted and labelled in the legend.}\label{fig:power.curves}
        \end{figure}

        Figure~\ref{fig:power.curves} provides a graphical representation of the averaged power results for each of the 40 tests when $\alpha=0.05$ and $n=200$. The power curves displayed within it were constructed in the following way.

        Recall that we considered 20 submodels and 20 specific cases per submodel. For each submodel, the 20 specific cases were assigned a number from 1 to 20 (according to their associated increasing value of $k$, $k_1$ or $k_2$ in Tables~\ref{table:symmetric.alternatives}--\ref{table:asymmetric.alternatives}). The smaller the number, the closer the specific case is to the Laplace distribution. For each test, and for each fixed value of $j=1,\ldots,20$, the powers against the $j$th specific case of the 20 submodels were averaged. In this way, we obtained, for each test, 20 averaged (mostly increasing) power values which we then connected to form a single power curve. In Figure~\ref{fig:power.curves}, each of the curves represents the averaged power values for a specific test. We have highlighted the power curves of the four tests with the best overall characteristics. The area under the solid black line -- divided by 20 -- is approximately 81.8\%, the average power of the $\text{DLO}_X$ test for $\alpha=0.05$ and $n=200$. An inspection of Figure~\ref{fig:power.curves} confirms that the range of values for $k$ (or $k_1$ or $k_2$) that we used indeed leads to the entire range of the power curve space being covered (see Section~\ref{sec:Design.of.the.MonteCarlo.study}).

        As expected, we observe in Figure~\ref{fig:power.curves} that all the curves are monotonic increasing for the submodels closest to the Laplace distribution (those numbered from 1 to 13). When the submodels are farther from the Laplace distribution, this distance is more difficult to measure consistently (for example for heavy-tailed or asymmetric distributions with extreme values). We therefore observe a small number of curves which are not monotonic.

   \subsubsection{The best tests against symmetric alternatives}\label{sec:symmetric.alternative}

        In some applications it will be reasonable to assume that the underlying distribution is either Laplace or some other symmetric distribution. Here we examine recommendations drawn from a consideration of the power results against the 12 symmetric submodels, including both heavy and light-tailed distributions. In Table~\ref{table:symmetric.average.power.n20.to.200} we present the average power results for the 10 best performing tests against ``Symmetric'' grouping of alternative distributions in Tables~\ref{new:table:alpha.0.05.n.20.specialized.ranks}--\ref{new:table:alpha.0.05.n.200.specialized.ranks}, together with the 10 best tests in terms of maximum and average gaps.

        \bgroup
        \def\arraystretch{1.0}
        \begin{table}[H]\centering
        \scriptsize
            \caption{The average \% power of the 10 best performing tests against the ``Symmetric'' group of alternatives extracted from Tables~\ref{new:table:alpha.0.05.n.20.specialized.ranks}--\ref{new:table:alpha.0.05.n.200.specialized.ranks}, as a function of sample size, $n$.}\label{table:symmetric.average.power.n20.to.200}
            \begin{center}
                \setlength\tabcolsep{4.87pt} 
\begin{tabular}{ccccccccccc}
                \hline
                $n = 20$ &  {\bf Wa} &  {\bf $\text{Me}_2^{(1)}$} & {\bf $\text{AP}_z$} & {\bf $\text{DLO}_X$} & {\bf $\text{Me}_{0.5}^{\scriptscriptstyle (2)}$} & {\bf Ku} & {\bf $\text{AP}_y$} & {\bf $\text{DLO}_Z$} & {\bf $\text{Ho}_U$} & {\bf $\text{AP}_e$}\\
                {\bf Power} &  44.5 &  43.5 & 42.9 & 42.8 & 42.8 & 42.4 & 40.6 & 40.3 & 39.5 & 39.1 \\
                {\bf Gap} &  - &  1.0 & 1.6 & 1.6 & 1.7 & 2.0 & 3.9 & 4.1 & 5.0 & 5.4 \\
                \hline
                $n = 50$ &  {\bf $\text{DLO}_Z$} & {\bf $\text{Ho}_U$} & {\bf $\text{DLO}_X$} & {\bf Wa} & {\bf GV} & {\bf $\text{Me}_{0.5}^{\scriptscriptstyle (2)}$} & {\bf BS} & {\bf $ \text{Me}_2^{\scriptscriptstyle (1)}$} & {\bf $\text{AP}_e$} & {\bf $\text{AP}_v$} \\
                {\bf Power} &  61.9 & 60.1 & 58.2 & 58.2 & 58.1 & 57.5 & 57.4 & 57.3 & 56.6 & 56.5\\
                {\bf Gap} &  - & 1.8 & 3.7 & 3.7 & 3.8 & 4.3 & 4.5 & 4.5 & 5.3 & 5.4 \\
                \hline
                $n = 100$ &  {\bf $\text{DLO}_Z$} & {\bf $\text{Ho}_U$} & {\bf GV} & {\bf BS} & {\bf $\text{DLO}_X$} & {\bf Wa} & {\bf $\text{Me}_{0.5}^{\scriptscriptstyle (2)}$} & {\bf $\text{Me}_2^{\scriptscriptstyle (1)}$} & {\bf LK} & {\bf Ku} \\
                {\bf Power} &  73.4 & 71.4 & 70.5 & 70.5 & 69.3 & 68.7 & 67.3 & 67.1 & 65.9 & 65.6 \\
\hline
                \end{tabular}
            \end{center}
        \end{table}
        \egroup

        From a consideration of Table~\ref{table:symmetric.average.power.n20.to.200}, the best tests against symmetric alternatives by sample size are the Wa test for $n=20$, and the $\text{DLO}_Z$ test for $n=50,100,200$. The best performing tests against symmetric alternatives regardless of sample size are the $\text{DLO}_X$ based test according to the maximum gap criterion, and the $\text{DLO}_Z$ based test according to the average gap criterion.

   \subsubsection{The best tests against symmetric heavy-tailed alternatives}\label{sec:heavy.alternative}

        Next we consider the scenario in which it can be assumed that the true underlying distribution is either Laplace or a heavier-tailed symmetric distribution. In Table~\ref{table:symmetric.heavy.average.power.n20.to.200}, we present the average power results for the 10 best performing tests against the ``Symmetric, heavy'' grouping of alternative distributions in Tables~\ref{new:table:alpha.0.05.n.20.specialized.ranks}--\ref{new:table:alpha.0.05.n.200.specialized.ranks} together with the 10 best tests in terms of maximum and average gaps.

        \bgroup
        \def\arraystretch{1.0}
        \begin{table}[H]\centering
        \scriptsize
            \caption{The average \% power of the 10 best performing tests against the ``Symmetric, heavy'' group of alternatives extracted from Tables~\ref{new:table:alpha.0.05.n.20.specialized.ranks}--\ref{new:table:alpha.0.05.n.200.specialized.ranks}, as a function of sample size, $n$.}\label{table:symmetric.heavy.average.power.n20.to.200}
            \begin{center}
                \setlength\tabcolsep{4.87pt} 
\begin{tabular}{ccccccccccc}
                \hline
                $n = 20$ &  {\bf BS} &  {\bf Wa} & {\bf $\text{AP}_z$} & {\bf Ku} & {\bf $\text{CK}_e$} & {\bf $\text{DLO}_X$} & {\bf  $\text{DLO}_Z$} & {\bf CvM} & {\bf $\text{AP}_a$} &     {\bf $\text{Me}_2^{\scriptscriptstyle (1)}$} \\
                {\bf Power} &  52.0 &  51.7 & 50.6 & 50.5 & 48.6 & 48.6 & 47.3 & 45.1 & 44.8 & 44.6 \\
                {\bf Gap} &  - &  0.3 & 1.5 & 1.6 & 3.4 & 3.4 & 4.7 & 6.9 & 7.2 & 7.4 \\
                \hline
                $n = 50$ &  {\bf $\text{AP}_a$} &   {\bf $\text{AP}_z$} &  {\bf $\text{AP}_y$} &  {\bf BS} & {\bf $\text{DLO}_Z$} & {\bf Wa} & {\bf $\text{DLO}_X$} & {\bf $\text{AP}_e$} &         {\bf Ku} & {\bf $\text{Me}_2^{\scriptscriptstyle (1)}$} \\
                {\bf Power} &  65.3 &  65.0 &  64.3 &  64.3 & 62.2 & 61.6 & 61.6 & 61.1 & 60.0 & 58.1 \\
                {\bf Gap} &  - &  0.3 &  1.0 &  1.0 & 3.1 & 3.7 & 3.7 & 4.2 & 5.3 & 7.2 \\
                \hline
                $n = 100$ &  {\bf $\text{AP}_y$} &  {\bf $\text{AP}_a$} & {\bf BS} & {\bf $\text{AP}_e$} & {\bf $\text{DLO}_Z$} & {\bf $\text{AP}_z$} & {\bf $\text{DLO}_X$} & {\bf $\text{AB}_{\text{KL}}$} & {\bf Wa} & {\bf $\text{AP}_v$} \\
                {\bf Power} &  72.4 &  71.6 & 71.1 & 70.8 & 69.4 & 68.9 & 68.1 & 67.7 & 67.5 & 66.5 \\
\hline
                \end{tabular}
            \end{center}
        \end{table}
        \egroup

        Considering the results in Table~\ref{table:symmetric.heavy.average.power.n20.to.200}, the best tests against symmetric heavy-tailed alternatives by sample size are, the BS based test for $n=20$, the $\text{AP}_a$ based test for $n=50$, and the $\text{AP}_y$ test for $n=100,200$. The best performing test against symmetric heavy-tailed alternatives regardless of sample size is, according to both the maximum gap and the average gap criteria, the $\text{BS}$ based test.

   \subsubsection{The best tests against symmetric light-tailed alternatives}\label{sec:light.alternative}

        Here we consider the scenario in which it can be assumed that the true underlying distribution is either Laplace or a lighter-tailed symmetric distribution. Table~\ref{table:symmetric.light.average.power.n20.to.200} present the average power results for the 10 best performing tests against the ``Symmetric, light'' grouping of alternative distributions in Tables~\ref{new:table:alpha.0.05.n.20.specialized.ranks}--\ref{new:table:alpha.0.05.n.200.specialized.ranks} together with the 10 best tests in terms of maximum and average gaps.

        \bgroup
        \def\arraystretch{1.0}
        \begin{table}[H]\centering
        \scriptsize
            \caption{The average \% power of the 10 best performing test against the ``Symmetric, light'' group of alternatives extracted from Tables~\ref{new:table:alpha.0.05.n.20.specialized.ranks}--\ref{new:table:alpha.0.05.n.200.specialized.ranks}, as a function of sample size, $n$.}\label{table:symmetric.light.average.power.n20.to.200}
            \begin{center}
                \setlength\tabcolsep{4.87pt} 
\begin{tabular}{ccccccccccc}
                \hline
                $n = 20$ &  {\bf $\text{CK}_c$} &  {\bf $\text{A}_{\text{rat}}$} &  {\bf $\text{CK}_v$} &  {\bf $\text{A}_{\text{ent}}$} &  {\bf AJ} &  {\bf $\text{AP}_y^{\scriptscriptstyle (\text{MLE})}$} & {\bf $\text{AP}_v$} & {\bf $\text{Me}_{0.5}^{\scriptscriptstyle (2)}$} & {\bf $\text{Ho}_K$} & {\bf $\text{Ho}_V$}\\
                {\bf Power} &  48.2 &  48.2 &  48.0 &  47.8 &  47.7 &  47.6 & 45.7 & 44.1 & 43.8 & 43.4 \\
                {\bf Gap} &  - &  0.1 &  0.3 &  0.5 &  0.5 &  0.6 & 2.5 & 4.1 & 4.5 & 4.8 \\
                \hline
                $n = 50$ &  {\bf $\text{A}_{\text{ent}}$} & {\bf AJ} & {\bf $\text{Ho}_U$} & {\bf GV} & {\bf $\text{AB}_{\text{TV}}$} & {\bf $\text{DLO}_Z$} & {\bf $\text{CK}_v$} &    {\bf $\text{A}_{\text{rat}}$} & {\bf $\text{CK}_c$} & {\bf $\text{AP}_y^{\scriptscriptstyle (\text{MLE})}$} \\
                {\bf Power} &  65.8 & 64.3 & 63.9 & 63.6 & 63.5 & 61.6 & 61.3 & 60.9 & 59.8 & 58.8\\
                {\bf Gap} &  - & 1.5 & 1.9 & 2.2 & 2.3 & 4.2 & 4.5 & 4.9 & 6.0 & 7.0 \\
                \hline
                $n = 100$ &  {\bf $\text{A}_{\text{ent}}$} &  {\bf $\text{Ho}_U$} &  {\bf $\text{DLO}_Z$} & {\bf GV} & {\bf $\text{AB}_{\text{TV}}$} & {\bf AJ} & {\bf $\text{DLO}_X$} &   {\bf $\text{Me}_{0.5}^{\scriptscriptstyle (2)}$} & {\bf Wa} & {\bf BS} \\
                {\bf Power} &  78.2 &  77.3 &  77.3 & 76.9 & 75.6 & 71.2 & 70.6 & 70.1 & 69.9 & 69.9 \\
\hline
                \end{tabular}
            \end{center}
        \end{table}
        \egroup

        From a consideration of the results in Table~\ref{table:symmetric.light.average.power.n20.to.200}, the best tests against symmetric light-tailed alternatives by sample size are, the $\text{CK}_c$ test for $n=20$, the $\text{A}_{\text{ent}}$ test for $n=50,100$, and the $\text{DLO}_Z$ test for $n=200$. The best performing test against symmetric light-tailed alternatives regardless of sample size is, according to both the maximum gap and the average gap criteria, the $\text{A}_{\text{ent}}$ based test.

   \subsubsection{The best tests against asymmetric alternatives}\label{sec:asymmetric.alternative}

        Finally, we consider the set-up in which it can be assumed that the true underlying distribution is either Laplace or some asymmetric distribution. Table~\ref{table:asymmetric.average.power.n20.to.200} presents the average power results for the ``Asymmetric'' grouping of alternative distributions in Tables~\ref{new:table:alpha.0.05.n.20.specialized.ranks}--\ref{new:table:alpha.0.05.n.200.specialized.ranks} together with the 10 best tests in terms of maximum and average gaps.

        \bgroup
        \def\arraystretch{1.0}
        \begin{table}[H]\centering
        \scriptsize
            \caption{The average \% power of the 10 best performing tests against the ``Asymmetric'' group of alternatives extracted from Tables~\ref{new:table:alpha.0.05.n.20.specialized.ranks}--\ref{new:table:alpha.0.05.n.200.specialized.ranks}, as a function of sample size, $n$.}\label{table:asymmetric.average.power.n20.to.200}
            \begin{center}
                \setlength\tabcolsep{4.87pt} 
\begin{tabular}{ccccccccccc}
                \hline
                $n = 20$ &  {\bf $\text{AP}_v$} &  {\bf $\text{AP}_y^{\scriptscriptstyle (\text{MLE})}$} &  {\bf $\text{AP}_e$} &  {\bf $\text{AP}_y$} & {\bf $\text{CK}_v$} & {\bf $\text{A}_{\text{rat}}$} & {\bf $\text{A}_{\text{ent}}$} & {\bf $\text{CK}_c$} & {\bf $\text{AP}_a$} & {\bf $\text{Z}_A$} \\
                {\bf Power} &  57.6 &  57.3 &  57.3 &  56.6 & 56.0 & 55.9 & 55.2 & 54.8 & 53.4 & 52.4 \\
                {\bf Gap} &  - &  0.2 &  0.2 &  0.9 & 1.5 & 1.6 & 2.4 & 2.8 & 4.1 & 5.2 \\
                \hline
                $n = 50$ &  {\bf $\text{A}_{\text{ent}}$} &  {\bf $\text{AP}_v$} & {\bf $\text{AP}_y^{\scriptscriptstyle (\text{MLE})}$} & {\bf $\text{CK}_v$} & {\bf $\text{A}_{\text{rat}}$} & {\bf AJ} & {\bf $\text{AP}_e$} & {\bf $\text{CK}_c$} & {\bf $\text{Z}_A$} & {\bf $\text{AP}_y$} \\
                {\bf Power} &  75.2 &  74.4 & 73.5 & 72.8 & 72.6 & 71.8 & 71.5 & 71.4 & 70.2 & 68.0 \\
                {\bf Gap} &  - &  0.9 & 1.7 & 2.4 & 2.7 & 3.5 & 3.7 & 3.8 & 5.0 & 7.2 \\
                \hline
                $n = 100$ &  {\bf $\text{A}_{\text{ent}}$} & {\bf $\text{Z}_A$} & {\bf LK} & {\bf $\text{AP}_v$} & {\bf $\text{DLO}_X$} & {\bf AJ} & {\bf $\text{AP}_y^{\scriptscriptstyle (\text{MLE})}$} & {\bf $\text{Z}_C$} & {\bf $\text{A}_{\text{rat}}$} & {\bf $\text{CK}_v$} \\
                {\bf Power} &  84.3 & 82.6 & 82.1 & 81.9 & 81.6 & 81.5 & 80.3 & 80.3 & 79.6 & 79.2 \\
\hline
                \end{tabular}
            \end{center}
        \end{table}
        \egroup

        The best tests against asymmetric alternatives by sample size are, the $\text{AP}_v$ test for $n=20$, the $\text{A}_{\text{ent}}$ test for $n=50,100$, and the $\text{LK}$ test for $n=200$. The best performing test against asymmetric alternatives regardless of sample size is, according to both the maximum gap and the average gap criteria, the $\text{A}_{\text{ent}}$ based test.

\section{Example}\label{sec:example}

\begin{figure}[H]\centering
        \vspace{-1.5cm}
        \includegraphics[width=0.6\textwidth]{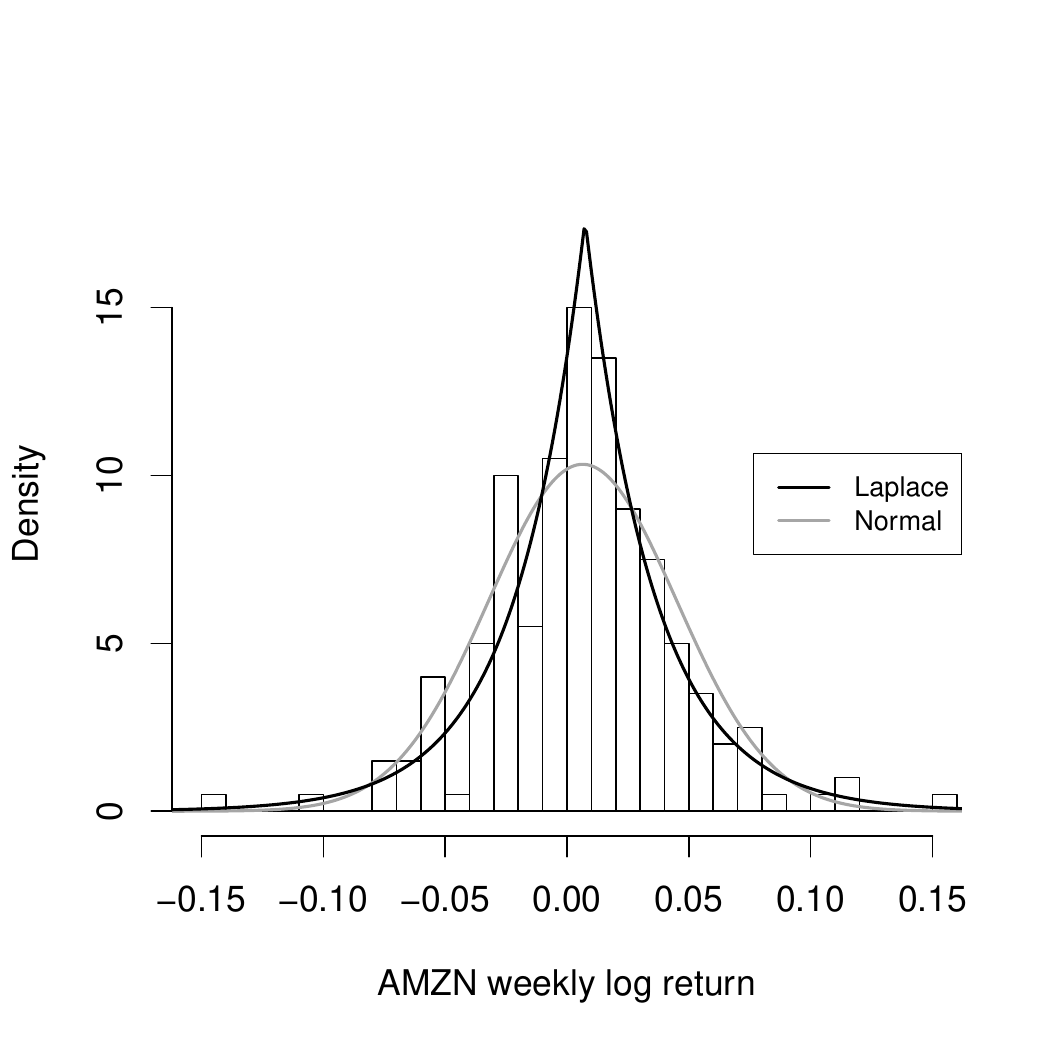}
         \vspace{-4mm}
         \caption{Histogram of $n=200$ Amazon.com Inc weekly log-returns with the maximum likelihood fitted Laplace and normal densities superimposed.}\label{fig:AMZN}
         \captionsetup{width=0.8\linewidth}
    \end{figure}

    It is well known that stock (log) returns often exhibit heavier tails than those of a normal distribution \citet[p.13]{Fan2015}, and they are sometimes modelled using a Laplace distribution \cite[see, e.g.,][]{Haas2006}.
    Here we present an analysis of a data set of $201$ closing prices $p_t$ ($t=1,\ldots,201$) of Amazon.com Inc (AMZN) stock on NASDAQ for each Friday during the period February 5, 2016 to December 13th, 2019. From them, we computed the $n=200$ log-returns $r_t=\log(p_{t+1}/p_{t})$, $t=1,\ldots,200$. We checked that the $r_t$'s are \textit{not} serially correlated, a property shared by most such series \citet[p.11]{Fan2015}.
    Figure~\ref{fig:AMZN} presents a histogram of the weekly log-returns, with the densities of the Laplace and normal distributions fitted to them using maximum likelihood. From a consideration of Figure~\ref{fig:AMZN}, it would appear that the underlying distribution could well be symmetric and potentially Laplace.

    \vspace{-2cm}

    In order to formally investigate whether the assumption of an underlying symmetric distribution is reasonable, we applied the characteristic function-based test of symmetry (around the mean) developed by \cite{Feuerverger1977} and implemented in the R package \texttt{symmetry} \citep{symmetry}. With a $p$-value of $0.9471$, the hypothesis of an underlying symmetric distribution is clearly not rejected. We thus applied the best Laplace test against symmetric alternatives for a sample size of $n=200$, namely that based on the $\textrm{DLO}_Z$ statistic (see Table~\ref{table:symmetric.average.power.n20.to.200}). The $p$-value of the test was $0.167$, from which we conclude that the Laplace distribution cannot be rejected as a potential underlying distribution for the weekly log-returns.

\section{Conclusion}\label{sec:conclusion}

    In this paper we have presented the results from a very extensive empirical power comparison of 40 goodness-of-fit tests for the univariate Laplace distribution. Previous comparative studies of tests for Laplaceness were all considerably narrower in scope and, as a consequence, likely introduced bias in terms of the selection of tests and alternative distributions considered. Here, we endeavoured to eliminate such selection bias by considering all the Laplace tests that we were able to identify and a wide range of alternative distributions including most of the models commonly considered in the literature when performing power studies.

    Our simulation study identified the $\text{DLO}_X$ based test as the best omnibus test, i.e., against all types of alternatives, whether symmetrical, heavy-tailed, light-tailed, or asymmetric. When it is reasonable to assume that the true underlying distribution is symmetric, we found that the $\text{DLO}_Z$ and $\text{DLO}_X$ based tests to be the most powerful. Globally, the most powerful tests against symmetric heavy-tailed and symmetric light-tailed alternatives were found to be the $\text{BS}$ and $\text{A}_{\text{ent}}$ based tests, respectively. The $\text{A}_{\text{ent}}$ based test was also identified as having the best global power performance against asymmetric alternatives.

\section*{Acknowledgements}


    This research includes computations performed using the computational cluster Katana supported by Research Technology Services at UNSW Sydney.
    Three anonymous referees have helped us greatly improve the paper. We are particularly grateful to one of the referees who spent a considerable amount of time improving its writing. We also thank the referees, the anonymous Associate Editor and the Editor in Chief for their patience.

%
%


\appendix

\section{Tables of results from the Monte Carlo study}\label{sec:power.tables.complete}

In this section, complete tables for the empirical power results can be found. Table~\ref{table:critical.values.A.1} shows the critical values for each of the 40 tests, each of the 4 sample sizes, and each of the 3 significance levels. For $\alpha=0.05$ and $n=20$, Table~\ref{new:table:alpha.0.05.n.20} shows the average powers for each of the 20 specified alternative distributions, for each test. Tables~\ref{new:table:alpha.0.05.n.50},~\ref{new:table:alpha.0.05.n.100} and~\ref{new:table:alpha.0.05.n.200} show the same results for $\alpha=0.05$ and $n=50, 100, 200$, respectively. For $\alpha=0.05$ and $n=20$, Table~\ref{new:table:alpha.0.05.n.20.specialized.ranks} shows the average powers against five groupings of alternative distributions: all models, symmetric, symmetric heavy-tailed, symmetric light-tailed, asymmetric. Tables~\ref{new:table:alpha.0.05.n.50.specialized.ranks},~\ref{new:table:alpha.0.05.n.100.specialized.ranks} and~\ref{new:table:alpha.0.05.n.200.specialized.ranks} show the same results for $\alpha=0.05$ and $n=50, 100, 200$, respectively. Similar tables for the significance levels of 0.01 and 0.10 are given in the supplementary materials document. The raw results of the 4,800 powers for each of the 40 tests are provided in the file \url{https://biostatisticien.eu/JSCS2022/Tables_Appendix_A_B.xlsx}.

\vspace{3mm}

    {
    \pagestyle{empty}
    \begin{landscape}
        \def\arraystretch{1.00}
        \begin{table}[htbp]\vspace{-1.51cm}
        \caption{Critical values as a function of sample size, $n$, and significance level, $\alpha$, for the 40 tests.\\For bilateral tests, {\normalfont (L)} and {\normalfont (R)} denote the left and right endpoint of the acceptance region, respectively.}
            \begin{center}
                \footnotesize
                \setlength\tabcolsep{2.00pt} 
                \vspace{-0.3cm}
\begin{tabular}{cccccccccccccccccc}
                \hline
                   {\bf n} & {\bf  $\alpha$} &   { AD} &  { CvM} &   { KS} &   { Ku} &   { Wa} &   { $\text{Z}_K$} &   { $\text{Z}_A$} &   { $\text{Z}_C$} & { $\text{DLO}_X$} & { $\text{DLO}_Z$}\hspace{0.3mm}(L) & { $\text{DLO}_Z$}\hspace{0.3mm}(R) &   { Ge} & { GV}\hspace{0.3mm}(L) & { GV}\hspace{0.3mm}(R) & { $\text{Ho}_K$}\hspace{0.3mm}(L) & { $\text{Ho}_K$}\hspace{0.3mm}(R) \\
                \hline
                        20 &       0.01 &      1.332 &      0.192 &      1.010 &      1.423 &      0.108 &      2.667 &      3.697 &     19.536 &      9.053 &     -2.697 &      2.557 &     11.138 &     -1.754 &      2.218 &      1.815 &     10.840 \\

                        20 &       0.05 &      0.923 &      0.130 &      0.867 &      1.259 &      0.080 &      1.941 &      3.545 &     12.969 &      5.930 &     -2.006 &      1.932 &      2.367 &     -1.534 &      1.334 &      2.049 &      8.342 \\

                        20 &       0.10 &      0.754 &      0.106 &      0.795 &      1.178 &      0.068 &      1.614 &      3.483 &     10.475 &      4.580 &     -1.663 &      1.619 &      1.044 &     -1.406 &      0.948 &      2.198 &      7.168 \\
                \hline
                        50 &       0.01 &      1.397 &      0.205 &      1.055 &      1.476 &      0.111 &      3.413 &      3.498 &     24.582 &      9.188 &     -2.578 &      2.613 &     20.959 &     -1.984 &      2.769 &      2.413 &     14.456 \\

                        50 &       0.05 &      0.961 &      0.139 &      0.903 &      1.305 &      0.083 &      2.496 &      3.425 &     16.365 &      5.978 &     -1.951 &      1.972 &      4.490 &     -1.677 &      1.749 &      2.692 &     10.211 \\

                        50 &       0.10 &      0.782 &      0.112 &      0.829 &      1.220 &      0.070 &      2.095 &      3.396 &     13.411 &      4.605 &     -1.635 &      1.651 &      2.030 &     -1.506 &      1.309 &      2.863 &      8.605 \\
\hline
                \end{tabular}
            \end{center}
            \label{table:critical.values.A.1}
            \normalsize
        \end{table}
    \end{landscape}
    }


    \newpage
   {
    \pagestyle{empty}
    \newgeometry{left=1.1in,right=1in,top=0.7in,bottom=0.7in,nohead}
    \begin{landscape}
        \def\arraystretch{1.00}
        \begin{table}[htbp]\vspace{-1.51cm}
               \caption{The average empirical \% power for each of the 40 tests, for $\alpha=0.05$ and $n=20$, against the 20 specified alternative distributions.\\ The ranges of the free parameter ($k$, $k_1$ or $k_2$) for each distribution are given in Tables~\ref{table:symmetric.alternatives} and~\ref{table:asymmetric.alternatives}. \\
               The highest powers (gap $\le 3.5\%$) are highlighted in grey and the largest is in bold.}
            \begin{center}
                \footnotesize
                \setlength\tabcolsep{2.00pt} 
                \vspace{-0.3cm}
\begin{tabular}{cccccccccccccccccccccc}
                    \hline
                    {\bf Type} & {\bf Alternative} &         AD &        CvM &         KS &         Ku &         Wa &        $\text{Z}_K$ &        $\text{Z}_A$ &        $\text{Z}_C$ &      $\text{DLO}_X$ &      $\text{DLO}_Z$ &         Ge &         GV &       $\text{Ho}_K$ &       $\text{Ho}_U$ &       $\text{Ho}_V$ &       $\text{Ho}_W$ &         LK &    $\text{A}_{\text{rat}}$ &  $\text{A}_{\text{ent}}$ &       $\text{AP}_v$ \\
                    \hline
                    \multirow{6}{*}{Sym., heavy} &   $\text{GED}(k<1)$ &       36.8 &       37.4 &       35.8 &      \cellcolor{gray!25} 41.8 &      \cellcolor{gray!25} 43.5 &       35.1 &       36.5 &       38.5 &      \cellcolor{gray!25} 43.4 &      \cellcolor{gray!25} 42.4 &       40.4 &       33.9 &       26.2 &       39.7 &       34.8 &       16.3 &       38.6 &       16.5 &       19.8 &       26.3 \\

                     &   $t(k<3.5)$ &       29.3 &       28.0 &       27.1 &       31.4 &       32.6 &       30.9 &       33.9 &      \cellcolor{gray!25} 36.2 &      \cellcolor{gray!25} 35.8 &       35.1 &     \cellcolor{gray!25}  39.0 &       34.3 &       32.0 &   \cellcolor{gray!25}    37.0 &       34.1 &       13.5 &       33.8 &       18.5 &       21.2 &       26.9 \\

                     & $\text{MixL}(w=0.5,k_1=0,k_2>1)$ &       71.3 &       76.9 &       74.7 &     \cellcolor{gray!25}  83.6 &    \cellcolor{gray!25}  {\bf 84.9} &       62.8 &       55.4 &       54.2 &       77.1 &       74.5 &       48.2 &       40.8 &       18.2 &       58.0 &       42.5 &       19.5 &       57.3 &       20.5 &       30.7 &       41.5 \\
\hline

            \end{tabular}
            \end{center}
            \label{new:table:alpha.0.05.n.20}
            \normalsize
        \end{table}
    \end{landscape}
    }

    \newpage    {
    \pagestyle{empty}
    \newgeometry{left=1.1in,right=1in,top=0.7in,bottom=0.7in,nohead}
    \begin{landscape}
        \def\arraystretch{1.00}
        \begin{table}[htbp]\vspace{-1.51cm}
               \caption{The average empirical \% power for each of the 40 tests, for $\alpha=0.05$ and $n=50$, against the 20 specified alternative distributions.\\ The ranges of the free parameter ($k$, $k_1$ or $k_2$) for each distribution are given in Tables~\ref{table:symmetric.alternatives} and~\ref{table:asymmetric.alternatives}. \\
               The highest powers (gap $\le 3.5\%$) are highlighted in grey and the largest is in bold.}
            \begin{center}
                \footnotesize
                \setlength\tabcolsep{2.00pt} 
                \vspace{-0.3cm}
\begin{tabular}{cccccccccccccccccccccc}
                    \hline
                {\bf Type} & {\bf Alternative} &         AD &        CvM &         KS &         Ku &         Wa &        $\text{Z}_K$ &        $\text{Z}_A$ &        $\text{Z}_C$ &      $\text{DLO}_X$ &      $\text{DLO}_Z$ &         Ge &         GV &       $\text{Ho}_K$ &       $\text{Ho}_U$ &       $\text{Ho}_V$ &       $\text{Ho}_W$ &         LK &    $\text{A}_{\text{rat}}$ &  $\text{A}_{\text{ent}}$ &       $\text{AP}_v$ \\
                \hline
                \multirow{6}{*}{Sym., heavy} &   $\text{GED}(k<1)$ &       50.3 &       51.3 &       49.3 &       56.2 &    \cellcolor{gray!25}   58.2 &       48.7 &       49.0 &       52.1 &   \cellcolor{gray!25}    59.2 &     \cellcolor{gray!25}  59.8 &       54.5 &       54.0 &       38.9 &       56.4 &       48.5 &       34.7 &       55.0 &       28.5 &       29.5 &       49.5 \\

                 &   $t(k<3.5)$ &       41.2 &       39.5 &       38.5 &       43.8 &       45.3 &       46.6 &       49.3 &      53.1 &       50.6 &       50.9 &      \cellcolor{gray!25} 56.5 &      \cellcolor{gray!25} 53.6 &       49.9 &      \cellcolor{gray!25} 54.2 &       51.5 &       43.0 &       48.8 &       25.1 &       31.9 &       45.3 \\

                 & $\text{MixL}(w=0.5,k_1=0,k_2>1)$ &       76.1 &       78.2 &       75.4 &     \cellcolor{gray!25}  83.0 &    \cellcolor{gray!25}   84.6 &       68.1 &       63.6 &       64.4 &     \cellcolor{gray!25}  83.3 &      \cellcolor{gray!25} 83.9 &       58.0 &       65.8 &       20.0 &       75.0 &       46.7 &       21.1 &       76.2 &       41.2 &       28.6 &       74.6 \\

                 & $\text{MixN}(w=0.5,k_1=0,k_2>3.5)$ &       74.4 &       78.5 &       76.5 &      \cellcolor{gray!25} 85.4 &     \cellcolor{gray!25} {\bf 86.9} &       63.2 &       41.1 &       38.7 &       73.6 &       75.6 &       18.2 &       37.0 &        1.7 &       45.7 &       12.3 &        3.4 &       59.2 &       47.2 &       15.3 &       72.5 \\
\hline
                \end{tabular}
            \end{center}
            \label{new:table:alpha.0.05.n.50}
             \normalsize
        \end{table}
    \end{landscape}
    }

    \newpage
    {
    \pagestyle{empty}
    \newgeometry{left=1.1in,right=1in,top=0.7in,bottom=0.7in,nohead}
    \begin{landscape}
        \def\arraystretch{1.00}
        \begin{table}[htbp]\vspace{-1.51cm}
               \caption{The average empirical \% power for each of the 40 tests, for $\alpha=0.05$ and $n=100$, against the 20 specified alternative distributions.\\ The ranges of the free parameter ($k$, $k_1$ or $k_2$) for each distribution are given in Tables~\ref{table:symmetric.alternatives} and~\ref{table:asymmetric.alternatives}. \\
               The highest powers (gap $\le 3.5\%$) are highlighted in grey and the largest is in bold.}
            \begin{center}
                \footnotesize
                \setlength\tabcolsep{2.00pt} 
                \vspace{-0.3cm}
\begin{tabular}{cccccccccccccccccccccc}
                    \hline
                {\bf Type} & {\bf Alternative} &         AD &        CvM &         KS &         Ku &         Wa &        $\text{Z}_K$ &        $\text{Z}_A$ &        $\text{Z}_C$ &      $\text{DLO}_X$ &      $\text{DLO}_Z$ &         Ge &         GV &       $\text{Ho}_K$ &       $\text{Ho}_U$ &       $\text{Ho}_V$ &       $\text{Ho}_W$ &         LK &    $\text{A}_{\text{rat}}$ &  $\text{A}_{\text{ent}}$ &       $\text{AP}_v$ \\
                \hline
                \multirow{6}{*}{Sym., heavy} &   $\text{GED}(k<1)$ &       59.9 &       61.1 &       59.2 &       65.8 &      \cellcolor{gray!25} 67.7 &       58.2 &       58.1 &       61.3 &   \cellcolor{gray!25}    68.9 &      \cellcolor{gray!25} {\bf 70.3} &       63.6 &       66.1 &       49.0 &       67.0 &       56.8 &       44.8 &       65.5 &       43.2 &       37.1 &       63.7 \\

                 &   $t(k<3.5)$ &       50.4 &       48.2 &       47.3 &       53.2 &       54.9 &       58.7 &       61.1 &       65.5 &       60.3 &       61.0 &     \cellcolor{gray!25}  68.9 &       65.8 &       63.4 &       65.9 &       63.9 &       58.9 &       59.2 &       33.3 &       41.0 &       60.5 \\

                 & $\text{MixL}(w=0.5,k_1=0,k_2>1)$ &       75.3 &       76.3 &       73.7 &     \cellcolor{gray!25}  80.7 &      \cellcolor{gray!25} 82.3 &       69.4 &       68.0 &       70.0 &     \cellcolor{gray!25}  82.8 &      \cellcolor{gray!25} {\bf 84.0} &       64.9 &       77.6 &       23.5 &       80.0 &       47.6 &       21.4 &       79.8 &       55.0 &       25.5 &       78.0 \\

                 & $\text{MixN}(w=0.5,k_1=0,k_2>3.5)$ &       76.2 &       78.2 &       76.0 &     \cellcolor{gray!25}  85.1 &      \cellcolor{gray!25} 86.8 &       63.4 &       42.3 &       41.5 &       75.1 &       78.4 &       12.8 &       47.2 &        0.6 &       51.8 &        5.1 &        1.7 &       68.5 &       67.6 &        6.2 &       80.4 \\
\hline
                \end{tabular}
            \end{center}
                \label{new:table:alpha.0.05.n.100}
            \normalsize
        \end{table}
    \end{landscape}
    }

    \newpage
    {
    \pagestyle{empty}
    \newgeometry{left=1.1in,right=1in,top=0.7in,bottom=0.7in,nohead}
    \begin{landscape}
        \def\arraystretch{1.00}
        \begin{table}[htbp]\vspace{-1.51cm}
               \caption{The average empirical \% power for each of the 40 tests, for $\alpha=0.05$ and $n=200$, against the 20 specified alternative distributions.\\ The ranges of the free parameter ($k$, $k_1$ or $k_2$) for each distribution are given in Tables~\ref{table:symmetric.alternatives} and~\ref{table:asymmetric.alternatives}. \\
               The highest powers (gap $\le 3.5\%$) are highlighted in grey and the largest is in bold.}
            \begin{center}
                \footnotesize
                \setlength\tabcolsep{2.00pt} 
                \vspace{-0.3cm}
\begin{tabular}{cccccccccccccccccccccc}
                    \hline
                {\bf Type} & {\bf Alternative} &         AD &        CvM &         KS &         Ku &         Wa &        $\text{Z}_K$ &        $\text{Z}_A$ &        $\text{Z}_C$ &      $\text{DLO}_X$ &      $\text{DLO}_Z$ &         Ge &         GV &       $\text{Ho}_K$ &       $\text{Ho}_U$ &       $\text{Ho}_V$ &       $\text{Ho}_W$ &         LK &    $\text{A}_{\text{rat}}$ &  $\text{A}_{\text{ent}}$ &       $\text{AP}_v$ \\
                \hline
                \multirow{6}{*}{Sym., heavy} &   $\text{GED}(k<1)$ &       68.8 &       69.9 &       68.1 &       74.0 &    \cellcolor{gray!25}   75.7 &       66.9 &       66.6 &       69.5 &      \cellcolor{gray!25} 76.9 &     \cellcolor{gray!25} {\bf 78.4} &       71.6 &      \cellcolor{gray!25} 75.5 &       58.6 &      \cellcolor{gray!25} 75.9 &       63.6 &       53.2 &       74.2 &       50.6 &       43.9 &       72.8 \\

                 &   $t(k<3.5)$ &       59.9 &       56.9 &       56.0 &       63.1 &       64.9 &       70.5 &       72.5 &       76.9 &       68.7 &       69.6 &     \cellcolor{gray!25}  80.0 &       76.1 &       76.0 &       76.2 &       75.3 &       72.5 &       68.4 &       37.3 &       53.5 &       74.7 \\

                 &  $\text{MixL}(w=0.5,k_1=0,k_2>1)$ &      73.6 &       74.2 &       71.7 &       78.3 &      \cellcolor{gray!25} 80.1 &       69.7 &       69.3 &       72.1 &      \cellcolor{gray!25} 81.4 &     \cellcolor{gray!25} {\bf  82.8} &       70.0 &      \cellcolor{gray!25} 80.3 &       29.5 &    \cellcolor{gray!25}   80.8 &       47.3 &       21.8 &      \cellcolor{gray!25} 79.4 &       50.7 &       21.1 & 77.0 \\
\hline
                \end{tabular}
            \end{center}
                \label{new:table:alpha.0.05.n.200}
            \normalsize
        \end{table}
    \end{landscape}
    }

    \newpage
    {
    \pagestyle{empty}
    \newgeometry{left=1.1in,right=1in,top=0.7in,bottom=0.7in,nohead}
    \begin{landscape}
        \def\arraystretch{1.00}
        \begin{table}[htbp]\vspace{-1.51cm}
               \caption{Ordered average \% empirical powers of the 40 tests, for $\alpha=0.05$ and $n=20$, against five groupings of alternative distributions: all models, symmetric, symmetric heavy-tailed, symmetric light-tailed, asymmetric.}
            \begin{center}
                \footnotesize
                \setlength\tabcolsep{2.00pt} 
                \vspace{-0.3cm}
\begin{tabular}{ccccccccccccccccccccc}
                    \hline
                        {\bf All models} &       $\text{AP}_y$ &       $\text{AP}_e$ &       $\text{AP}_z$ &       $\text{AP}_v$ &       $\text{CK}_v$ &   $\text{AP}_y^{\scriptscriptstyle (\text{MLE})}$ &     $\text{Me}_2^{\scriptscriptstyle (1)}$ &       $\text{AP}_a$ &         Wa &      $\text{DLO}_X$ &   $\text{Me}_{0.5}^{\scriptscriptstyle (2)}$ &  $\text{A}_{\text{ent}}$ &       $\text{CK}_c$ &         Ku &         LK &    $\text{A}_{\text{rat}}$ &         AD &        $\text{SR}^{\star}$ &        $\text{Z}_K$ &        $\text{Z}_A$ \\
                    \hline
                    Power &       47.0 &       46.4 &       46.2 &       45.3 &       45.2 &       44.9 &       44.7 &       44.5 &       44.2 &       44.0 &       43.7 &       42.9 &       42.7 &       42.1 &       42.0 &       41.9 &       41.1 &       40.8 &       40.7 &       40.5 \\

                     Gap &        0.0 &        0.6 &        0.8 &        1.7 &        1.8 &        2.1 &        2.3 &        2.5 &        2.8 &        3.0 &        3.3 &        4.1 &        4.3 &        4.9 &        5.0 &        5.1 &        5.9 &        6.2 &        6.3 &        6.5 \\

                    Rank &          1 &          2 &          3 &          4 &          5 &          6 &          7 &          8 &          9 &         10 &         11 &         12 &         13 &         14 &         15 &         16 &         17 &         18 &         19 &         20 \\
                    \hline
                         &        CvM &        $\text{Z}_C$ &    $\text{AB}_{\text{KL}}$ &       $\text{CK}_e$ &    $\text{AB}_{\text{He}}$ &         AJ &         KS &    $\text{AB}_{\text{Je}}$ &         BS &       $\text{Ho}_U$ &         GV &       $\text{Ho}_V$ &      $\text{DLO}_Z$ &         KP &       $\text{Ho}_K$ &       $\text{Ho}_W$ &    $\text{AB}_{\chi^2}$ &         Ge &    $\text{AB}_{\text{TV}}$ &         SD \\
\end{tabular}
            \end{center}
            \label{new:table:alpha.0.05.n.20.specialized.ranks}
            \normalsize
        \end{table}
    \end{landscape}
    }

    \newpage
    {
    \pagestyle{empty}
    \newgeometry{left=1.1in,right=1in,top=0.7in,bottom=0.7in,nohead}
    \begin{landscape}
        \def\arraystretch{1.00}
        \begin{table}[htbp]\vspace{-1.51cm}
               \caption{Ordered average \% empirical powers of the 40 tests, for $\alpha=0.05$ and $n=50$, against five groupings of alternative distributions: all models, symmetric, symmetric heavy-tailed, symmetric light-tailed, asymmetric.}
            \begin{center}
                \footnotesize
                \setlength\tabcolsep{2.00pt} 
                \vspace{-0.3cm}
\begin{tabular}{ccccccccccccccccccccc}
                    \hline
                    {\bf All models} &       $\text{AP}_v$ &       $\text{AP}_e$ &   $\text{AP}_y^{\scriptscriptstyle (\text{MLE})}$ &      $\text{DLO}_X$ &       $\text{AP}_y$ &   $\text{Me}_{0.5}^{\scriptscriptstyle (2)}$ &     $\text{Me}_2^{\scriptscriptstyle (1)}$ &         LK &         Wa &       $\text{CK}_v$ &       $\text{CK}_c$ &  $\text{A}_{\text{ent}}$ &         AJ &    $\text{A}_{\text{rat}}$ &    $\text{AB}_{\text{He}}$ &        $\text{Z}_A$ &    $\text{AB}_{\text{Je}}$ &         Ku &        $\text{Z}_C$ &    $\text{AB}_{\text{KL}}$ \\
                \hline
                Power &       63.7 &       62.6 &       62.3 &       60.8 &       60.4 &       60.4 &       60.1 &       59.0 &       58.9 &       58.8 &       58.2 &       58.1 &       58.1 &       56.9 &       56.7 &       55.9 &       55.4 &       55.3 &       54.6 &       54.0 \\

                 Gap &        0.0 &        1.1 &        1.4 &        2.8 &        3.2 &        3.3 &        3.5 &        4.7 &        4.7 &        4.8 &        5.4 &        5.5 &        5.6 &        6.8 &        7.0 &        7.8 &        8.2 &        8.3 &        9.1 &        9.6 \\

                Rank &          1 &          2 &          3 &          4 &          5 &          6 &          7 &          8 &          9 &         10 &         11 &         12 &         13 &         14 &         15 &         16 &         17 &         18 &         19 &         20 \\
                \hline
                    {\bf } &         AD &        $\text{SR}^{\star}$ &         GV &        $\text{Z}_K$ &      $\text{DLO}_Z$ &       $\text{AP}_a$ &        CvM &       $\text{Ho}_U$ &         BS &         KS &       $\text{Ho}_V$ &       $\text{CK}_e$ &       $\text{AP}_z$ &    $\text{AB}_{\text{TV}}$ &       $\text{Ho}_K$ &       $\text{Ho}_W$ &         KP &    $\text{AB}_{\chi^2}$ &         Ge &         SD \\
\end{tabular}
            \end{center}
            \label{new:table:alpha.0.05.n.50.specialized.ranks}
            \normalsize
        \end{table}
    \end{landscape}
    }

    \newpage
    {
    \pagestyle{empty}
    \newgeometry{left=1.1in,right=1in,top=0.7in,bottom=0.7in,nohead}
    \begin{landscape}
        \def\arraystretch{1.00}
        \begin{table}[htbp]\vspace{-1.51cm}
               \caption{Ordered average \% empirical powers of the 40 tests, for $\alpha=0.05$ and $n=100$, against five groupings of alternative distributions: all models, symmetric, symmetric heavy-tailed, symmetric light-tailed, asymmetric.}
            \begin{center}
                \footnotesize
                \setlength\tabcolsep{2.00pt} 
                \vspace{-0.3cm}
\begin{tabular}{ccccccccccccccccccccc}
                    \hline
                        {\bf All models} &      $\text{DLO}_X$ &         LK &         Wa &       $\text{AP}_v$ &   $\text{Me}_{0.5}^{\scriptscriptstyle (2)}$ &     $\text{Me}_2^{\scriptscriptstyle (1)}$ &   $\text{AP}_y^{\scriptscriptstyle (\text{MLE})}$ &       $\text{AP}_e$ &    $\text{AB}_{\text{He}}$ &         Ku &         AJ &         GV &        $\text{Z}_A$ &    $\text{AB}_{\text{Je}}$ &  $\text{A}_{\text{ent}}$ &      $\text{DLO}_Z$ &        $\text{Z}_C$ &    $\text{A}_{\text{rat}}$ &       $\text{Ho}_U$ &       $\text{AP}_y$ \\
                    \hline
                    Power &       74.3 &       72.3 &       71.8 &       71.7 &       71.7 &       71.4 &       69.0 &       68.6 &       68.1 &       67.8 &       67.8 &       67.5 &       67.4 &       66.9 &       66.7 &       66.6 &       66.5 &       65.4 &       65.1 &       65.0 \\

                     Gap &        0.0 &        1.9 &        2.4 &        2.5 &        2.6 &        2.8 &        5.2 &        5.7 &        6.2 &        6.5 &        6.5 &        6.8 &        6.8 &        7.4 &        7.6 &        7.6 &        7.8 &        8.8 &        9.2 &        9.2 \\

                    Rank &          1 &          2 &          3 &          4 &          5 &          6 &          7 &          8 &          9 &         10 &         11 &         12 &         13 &         14 &         15 &         16 &         17 &         18 &         19 &         20 \\
                    \hline
                        {\bf } &    $\text{AB}_{\text{KL}}$ &       $\text{CK}_v$ &         BS &    $\text{AB}_{\text{TV}}$ &         AD &        $\text{SR}^{\star}$ &        CvM &       $\text{CK}_c$ &        $\text{Z}_K$ &         KS &       $\text{Ho}_V$ &       $\text{Ho}_K$ &       $\text{AP}_a$ &       $\text{CK}_e$ &       $\text{Ho}_W$ &       $\text{AP}_z$ &         KP &    $\text{AB}_{\chi^2}$ &         Ge &         SD \\
\end{tabular}
            \end{center}
            \label{new:table:alpha.0.05.n.100.specialized.ranks}
            \normalsize
        \end{table}
    \end{landscape}
    }

    \newpage
    {
    \pagestyle{empty}
    \newgeometry{left=1.1in,right=1in,top=0.7in,bottom=0.7in,nohead}
    \begin{landscape}
        \def\arraystretch{1.00}
        \begin{table}[htbp]\vspace{-1.51cm}
               \caption{Ordered average \% empirical powers of the 40 tests, for $\alpha=0.05$ and $n=200$, against five groupings of alternative distributions: all models, symmetric, symmetric heavy-tailed, symmetric light-tailed, asymmetric.}
            \begin{center}
                \footnotesize
                \setlength\tabcolsep{2.00pt} 
                \vspace{-0.3cm}
\begin{tabular}{ccccccccccccccccccccc}
                \hline
                    {\bf All models} &      $\text{DLO}_X$ &         Wa &         LK &     $\text{Me}_2^{\scriptscriptstyle (1)}$ &   $\text{Me}_{0.5}^{\scriptscriptstyle (2)}$ &         Ku &    $\text{AB}_{\text{He}}$ &         BS &    $\text{AB}_{\text{Je}}$ &       $\text{AP}_v$ &      $\text{DLO}_Z$ &         GV &        $\text{Z}_C$ &        $\text{Z}_A$ &    $\text{AB}_{\text{KL}}$ &       $\text{Ho}_U$ &   $\text{AP}_y^{\scriptscriptstyle (\text{MLE})}$ &       $\text{AP}_e$ &         AD &        CvM \\
                \hline
                Power &       81.8 &       80.7 &       80.3 &       79.5 &       79.1 &       78.0 &       77.3 &       77.1 &       76.8 &       76.8 &       76.3 &       75.3 &       75.3 &       75.1 &       74.7 &       74.7 &       74.2 &       73.9 &       73.7 &       73.6 \\

                 Gap &        0.0 &        1.1 &        1.5 &        2.3 &        2.7 &        3.8 &        4.5 &        4.7 &        5.0 &        5.0 &        5.5 &        6.5 &        6.5 &        6.7 &        7.1 &        7.1 &        7.7 &        7.9 &        8.2 &        8.2 \\

                Rank &          1 &          2 &          3 &          4 &          5 &          6 &          7 &          8 &          9 &         10 &         11 &         12 &         13 &         14 &         15 &         16 &         17 &         18 &         19 &         20 \\
                \hline
                    {\bf } &    $\text{AB}_{\text{TV}}$ &        $\text{SR}^{\star}$ &         AJ &       $\text{AP}_y$ &  $\text{A}_{\text{ent}}$ &        $\text{Z}_K$ &         KS &    $\text{A}_{\text{rat}}$ &       $\text{CK}_v$ &       $\text{CK}_c$ &       $\text{Ho}_K$ &       $\text{Ho}_V$ &       $\text{AP}_a$ &       $\text{CK}_e$ &       $\text{Ho}_W$ &       $\text{AP}_z$ &         KP &         Ge &    $\text{AB}_{\chi^2}$ &         SD \\
\end{tabular}
            \end{center}
            \label{new:table:alpha.0.05.n.200.specialized.ranks}
            \normalsize
        \end{table}
    \end{landscape}
    }


    \newpage

   {
    \pagestyle{empty}
    \newgeometry{left=1.1in,right=1in,top=0.7in,bottom=0.7in,nohead}
    \begin{landscape}
        \def\arraystretch{1.00}
        \begin{table}[htbp]\vspace{-1.51cm}
        \vspace{-7mm}
        \textbf{{\Large Supplementary materials document}\\[1mm] for ``A comprehensive empirical power comparison of univariate goodness-of-fit tests for the Laplace distribution''} \\
               \caption{The average empirical \% power for each of the 40 tests, for $\alpha=0.01$ and $n=20$, against the 20 specified alternative distributions.\\ The ranges of the free parameter ($k$, $k_1$ or $k_2$) for each distribution are given in Tables~\ref{table:symmetric.alternatives} and~\ref{table:asymmetric.alternatives}. \\
               The highest powers (gap $\le 3.5\%$) are highlighted in grey and the largest is in bold.}

            \begin{center}
                \footnotesize
                \setlength\tabcolsep{2.00pt} 
    \vspace{-0.3cm}
\begin{tabular}{cccccccccccccccccccccc}
    \toprule
    {\bf Type} & {\bf Alternative} &         AD &        CvM &         KS &         Ku &         Wa &        $\text{Z}_K$ &        $\text{Z}_A$ &        $\text{Z}_C$ &      $\text{DLO}_X$ &      $\text{DLO}_Z$ &         Ge &         GV &       $\text{Ho}_K$ &       $\text{Ho}_U$ &       $\text{Ho}_V$ &       $\text{Ho}_W$ &         LK &    $\text{A}_{\text{rat}}$ &  $\text{A}_{\text{ent}}$ &       $\text{AP}_v$ \\
    \midrule
     \multirow{6}{*}{Sym., heavy} & GED($k<1$)   &  24.3 & 25.7 & 24.5 & \cellcolor{gray!25} 32 & \cellcolor{gray!25} {\bf 34.2} & 22.9 & 23.3 & 27.9 & \cellcolor{gray!25} 32.7 & \cellcolor{gray!25} 31.5 & 26.7 & 21.4 & 15.3 & 27.7 & 24.1 & 7.2 & 28.1 & 10.9 & 13.4 & 18.8 \\
      & $t(k<3.5)$   &  20.1 & 19.7 & 18.7 & 24 & \cellcolor{gray!25} 25.4 & 21.6 & 23.4 & \cellcolor{gray!25} 28 & \cellcolor{gray!25} 27.6 & \cellcolor{gray!25} 26.7 & \cellcolor{gray!25} {\bf 28.6} & 24 & 21.6 & \cellcolor{gray!25} 27.6 & 25 & 5.8 & \cellcolor{gray!25} 26.3 & 12.6 & 14.2 & 19.5 \\
      & MixL($w=0.5,k_1=0,k_2>1$)   &  47.9 & 58 & 56.9 & \cellcolor{gray!25} 73.7 & \cellcolor{gray!25} {\bf 75.5} & 39.9 & 29.2 & 33.8 & 59.6 & 56.5 & 24.6 & 21.3 & 7 & 33.1 & 24.5 & 7.9 & 37.6 & 9.9 & 16.3 & 25.1 \\
      & MixN($w=0.5,k_1=0,k_2>3.5$)   &  37.4 & 48.4 & 49.8 & \cellcolor{gray!25} 67.9 & \cellcolor{gray!25} {\bf 68.2} & 31.6 & 16.4 & 16.8 & 39.1 & 37.8 & 8 & 9.8 & 1.3 & 14 & 9.6 & 3.5 & 23 & 5.1 & 8.4 & 12.7 \\
      & NIG($k_1<0.6,k_2=0$)   &  4.8 & 4.5 & 4.4 & 7.3 & \cellcolor{gray!25} 8.3 & 6 & 7.2 & \cellcolor{gray!25} 10.6 & \cellcolor{gray!25} 9.8 & \cellcolor{gray!25} 9.1 & \cellcolor{gray!25} {\bf 11.2} & \cellcolor{gray!25} 8.5 & 7.3 & \cellcolor{gray!25} 9.8 & \cellcolor{gray!25} 8.6 & 3.4 & \cellcolor{gray!25} 9.1 & 2 & 2.3 & 4.6 \\
\bottomrule
    \end{tabular}%
            \end{center}
            \label{new:table:alpha.0.01.n.20}
            \normalsize
        \end{table}
    \end{landscape}
    }


        \newpage
   {
    \pagestyle{empty}
    \newgeometry{left=1.1in,right=1in,top=0.7in,bottom=0.7in,nohead}
    \begin{landscape}
        \def\arraystretch{1.00}
        \begin{table}[htbp]\vspace{-1.51cm}
               \caption{The average empirical \% power for each of the 40 tests, for $\alpha=0.01$ and $n=50$, against the 20 specified alternative distributions.\\ The ranges of the free parameter ($k$, $k_1$ or $k_2$) for each distribution are given in Tables~\ref{table:symmetric.alternatives} and~\ref{table:asymmetric.alternatives}. \\
               The highest powers (gap $\le 3.5\%$) are highlighted in grey and the largest is in bold.}

            \begin{center}
                \footnotesize
                \setlength\tabcolsep{2.00pt} 
    \vspace{-0.3cm}
\begin{tabular}{cccccccccccccccccccccc}
    \toprule
    {\bf Type} & {\bf Alternative} &         AD &        CvM &         KS &         Ku &         Wa &        $\text{Z}_K$ &        $\text{Z}_A$ &        $\text{Z}_C$ &      $\text{DLO}_X$ &      $\text{DLO}_Z$ &         Ge &         GV &       $\text{Ho}_K$ &       $\text{Ho}_U$ &       $\text{Ho}_V$ &       $\text{Ho}_W$ &         LK &    $\text{A}_{\text{rat}}$ &  $\text{A}_{\text{ent}}$ &       $\text{AP}_v$ \\
    \midrule
     \multirow{6}{*}{Sym., heavy} & GED($k<1$)   &  38.7 & 40.5 & 38.6 & \cellcolor{gray!25} 47.4 & \cellcolor{gray!25} 50.2 & 36.9 & 36.6 & 42.5 & \cellcolor{gray!25} 50.3 & \cellcolor{gray!25} {\bf 50.6} & 41 & 42.5 & 25.1 & 45.4 & 37.9 & 21.4 & 45.7 & 23.3 & 23.9 & 41.3 \\
      & $t(k<3.5)$   &  31.9 & 31.1 & 29.8 & 36.1 & 38 & 36.9 & 38.2 & \cellcolor{gray!25} 45.1 & \cellcolor{gray!25} 43.1 & \cellcolor{gray!25} 43 & \cellcolor{gray!25} {\bf 46} & \cellcolor{gray!25} 44.6 & 37.8 & \cellcolor{gray!25} 45.3 & 42.4 & 25.2 & 41.8 & 20.9 & 24.6 & 37.9 \\
      & MixL($w=0.5,k_1=0,k_2>1$)   &  59.7 & 65.6 & 61.7 & \cellcolor{gray!25} 76 & \cellcolor{gray!25} {\bf 78.8} & 47 & 36.5 & 43 & 73.9 & 74.4 & 29.8 & 42.9 & 6.3 & 52.8 & 26.1 & 8.6 & 61 & 26.4 & 15 & 63.2 \\
      & MixN($w=0.5,k_1=0,k_2>3.5$)   &  53.2 & 62.2 & 61 & \cellcolor{gray!25} 77.5 & \cellcolor{gray!25} {\bf 79.7} & 39.7 & 15.4 & 15.4 & 55.9 & 58 & 3.7 & 14.4 & 0.2 & 18.5 & 3.5 & 0.6 & 37.8 & 28 & 6.1 & 57 \\
      & NIG($k_1<0.7,k_2=0$)   &  8.6 & 8.2 & 7.9 & 12.5 & 14.2 & 11.7 & 12.1 & \cellcolor{gray!25} 18 & \cellcolor{gray!25} 17.6 & \cellcolor{gray!25} 17.7 & \cellcolor{gray!25} {\bf 18.6} & \cellcolor{gray!25} 17.7 & 11.7 & \cellcolor{gray!25} 18.4 & \cellcolor{gray!25} 15.6 & 10.1 & \cellcolor{gray!25} 16.3 & 2.5 & 3.6 & 11.6 \\
\bottomrule
    \end{tabular}%
            \end{center}
            \label{new:table:alpha.0.01.n.50}
            \normalsize
        \end{table}
    \end{landscape}
    }



        \newpage
   {
    \pagestyle{empty}
    \newgeometry{left=1.1in,right=1in,top=0.7in,bottom=0.7in,nohead}
    \begin{landscape}
        \def\arraystretch{1.00}
        \begin{table}[htbp]\vspace{-1.51cm}
               \caption{The average empirical \% power for each of the 40 tests, for $\alpha=0.01$ and $n=100$, against the 20 specified alternative distributions.\\ The ranges of the free parameter ($k$, $k_1$ or $k_2$) for each distribution are given in Tables~\ref{table:symmetric.alternatives} and~\ref{table:asymmetric.alternatives}. \\
               The highest powers (gap $\le 3.5\%$) are highlighted in grey and the largest is in bold.}

            \begin{center}
                \footnotesize
                \setlength\tabcolsep{2.00pt} 
    \vspace{-0.3cm}
\begin{tabular}{cccccccccccccccccccccc}
    \toprule
    {\bf Type} & {\bf Alternative} &         AD &        CvM &         KS &         Ku &         Wa &        $\text{Z}_K$ &        $\text{Z}_A$ &        $\text{Z}_C$ &      $\text{DLO}_X$ &      $\text{DLO}_Z$ &         Ge &         GV &       $\text{Ho}_K$ &       $\text{Ho}_U$ &       $\text{Ho}_V$ &       $\text{Ho}_W$ &         LK &    $\text{A}_{\text{rat}}$ &  $\text{A}_{\text{ent}}$ &       $\text{AP}_v$ \\
    \midrule
     \multirow{6}{*}{Sym., heavy} & GED($k<1$)   &  49.5 & 51.4 & 49.3 & 58 & \cellcolor{gray!25} 60.6 & 47.4 & 47.1 & 52.9 & \cellcolor{gray!25} 61.4 & \cellcolor{gray!25} {\bf 62.5} & 51.2 & 56.6 & 35.1 & 57.8 & 46.7 & 32.5 & 57.4 & 37.6 & 32.1 & 56.2 \\
      & $t(k<3.5)$   &  40.9 & 39.7 & 38.4 & 45.2 & 47.1 & 49.2 & 50.2 & \cellcolor{gray!25} 58.3 & 53.5 & 54 & \cellcolor{gray!25} {\bf 59.2} & \cellcolor{gray!25} 58.1 & 52.1 & \cellcolor{gray!25} 58.3 & 55.2 & 47.7 & 52.8 & 29.7 & 33.1 & 53.4 \\
      & MixL($w=0.5,k_1=0,k_2>1$)   &  62.9 & 66 & 61.9 & \cellcolor{gray!25} 74.2 & \cellcolor{gray!25} 77 & 51.1 & 46 & 52.6 & \cellcolor{gray!25} 76 & \cellcolor{gray!25} {\bf 77.4} & 35.9 & 62 & 7.1 & 66.4 & 26.1 & 8.3 & 70.6 & 44.1 & 13.6 & 70.1 \\
      & MixN($w=0.5,k_1=0,k_2>3.5$)   &  58 & 63.5 & 61.1 & \cellcolor{gray!25} 77.1 & \cellcolor{gray!25} {\bf 79.7} & 39.9 & 15.2 & 15.3 & 60.2 & 64 & 1.5 & 20.5 & 0 & 23.5 & 0.8 & 0.2 & 48.6 & 52.3 & 1.9 & 68.6 \\
      & NIG($k_1<0.7,k_2=0$)   &  14.3 & 13.8 & 13.1 & 19.5 & 21.5 & 19 & 18.9 & \cellcolor{gray!25} 26.7 & \cellcolor{gray!25} 26 & \cellcolor{gray!25} 26.8 & \cellcolor{gray!25} 27.3 & \cellcolor{gray!25} 28.3 & 17.8 & \cellcolor{gray!25} {\bf 28.6} & 22.9 & 16.7 & 24.7 & 5.4 & 6.2 & 22.6 \\
\bottomrule
    \end{tabular}%
            \end{center}
            \label{new:table:alpha.0.01.n.100}
            \normalsize
        \end{table}
    \end{landscape}
    }



        \newpage
   {
    \pagestyle{empty}
    \newgeometry{left=1.1in,right=1in,top=0.7in,bottom=0.7in,nohead}
    \begin{landscape}
        \def\arraystretch{1.00}
        \begin{table}[htbp]\vspace{-1.51cm}
               \caption{The average empirical \% power for each of the 40 tests, for $\alpha=0.01$ and $n=200$, against the 20 specified alternative distributions.\\ The ranges of the free parameter ($k$, $k_1$ or $k_2$) for each distribution are given in Tables~\ref{table:symmetric.alternatives} and~\ref{table:asymmetric.alternatives}. \\
               The highest powers (gap $\le 3.5\%$) are highlighted in grey and the largest is in bold.}

            \begin{center}
                \footnotesize
                \setlength\tabcolsep{2.00pt} 
    \vspace{-0.3cm}
\begin{tabular}{cccccccccccccccccccccc}
    \toprule
    {\bf Type} & {\bf Alternative} &         AD &        CvM &         KS &         Ku &         Wa &        $\text{Z}_K$ &        $\text{Z}_A$ &        $\text{Z}_C$ &      $\text{DLO}_X$ &      $\text{DLO}_Z$ &         Ge &         GV &       $\text{Ho}_K$ &       $\text{Ho}_U$ &       $\text{Ho}_V$ &       $\text{Ho}_W$ &         LK &    $\text{A}_{\text{rat}}$ &  $\text{A}_{\text{ent}}$ &       $\text{AP}_v$ \\
    \midrule
     \multirow{6}{*}{Sym., heavy} & GED($k<1$)   &  59.7 & 61.6 & 59.5 & 67.5 & \cellcolor{gray!25} 69.9 & 57.4 & 57.4 & 62.3 & \cellcolor{gray!25} 70.9 & \cellcolor{gray!25} {\bf 72.2} & 60.4 & 67.9 & 45.2 & 68.3 & 54.2 & 41.7 & 67.5 & 45.6 & 39.8 & 66.5 \\
      & $t(k<3.5)$   &  50.1 & 48.1 & 47.1 & 54.4 & 56.3 & 61.6 & 62.7 & \cellcolor{gray!25} 70.8 & 62.7 & 63.4 & \cellcolor{gray!25} {\bf 71.7} & \cellcolor{gray!25} 69.8 & 66 & \cellcolor{gray!25} 69.9 & 67.5 & 63 & 62.8 & 34.8 & 43.1 & 68 \\
      & MixL($w=0.5,k_1=0,k_2>3.5$)   &  62.9 & 64.7 & 61 & 72 & \cellcolor{gray!25} 74.8 & 54.2 & 53.6 & 59.5 & \cellcolor{gray!25} 75.4 & \cellcolor{gray!25} {\bf 76.9} & 43.5 & 70.9 & 8.7 & 71.8 & 25.5 & 8.3 & 72.3 & 41.2 & 11.2 & 70 \\
      & MixN($w=0.5,k_1=0,k_2>3.5$)   &  59.8 & 62.9 & 60.2 & \cellcolor{gray!25} 77 & \cellcolor{gray!25} {\bf 80} & 38.4 & 15.4 & 16.2 & 60.3 & 64.9 & 0.5 & 25.7 & 0 & 27.2 & 0.1 & 0.3 & 54.6 & 56.3 & 0.2 & 72.1 \\
      & NIG($k_1<0.65,k_2=0$)   &  23.5 & 22.7 & 21.7 & 29.9 & 32.3 & 30.3 & 30 & \cellcolor{gray!25} 39.3 & 37.3 & 38.7 & \cellcolor{gray!25} 39.7 & \cellcolor{gray!25} 42.4 & 27.5 & \cellcolor{gray!25} {\bf 42.6} & 32.9 & 25.2 & 36.5 & 9 & 11.5 & 37.3 \\
\bottomrule
    \end{tabular}%
            \end{center}
            \label{new:table:alpha.0.01.n.200}
            \normalsize
        \end{table}
    \end{landscape}
    }



\newpage
    {
    \pagestyle{empty}
    \newgeometry{left=1.1in,right=1in,top=0.7in,bottom=0.7in,nohead}
    \begin{landscape}
        \def\arraystretch{1.00}
        \begin{table}[htbp]\vspace{-1.51cm}
               \caption{Ordered average \% empirical powers of the 40 tests, for $\alpha=0.01$ and $n=20$, against five groupings of alternative distributions: all models, symmetric, symmetric heavy-tailed, symmetric light-tailed, asymmetric.}
            \begin{center}
                \footnotesize
                \setlength\tabcolsep{2.00pt} 
    \vspace{-0.3cm}
\begin{tabular}{lcccccccccccccccccccc}
    \toprule
    \multicolumn{1}{c}{\textbf{All models}} & $\text{AP}_z$ & $\text{AP}_y$ & $\text{Wa}$ & $\text{AP}_e$ & $\text{ME}^{((1)}_2$ & $\text{AP}_v$ & $\text{AP}^{(\text{MLE})}_y$ & $\text{AP}_a$ & $\text{CK}_v$   & $\text{DLO}_X$ & $\text{ME}^{(2)}_{0.5}$ & $\text{Ku}$ & $\text{CK}_e$   & $\text{LK}$ & $\text{Z}_K$ & $\text{A}_{\text{ent}}$ & $\text{CK}_c$   & $\text{AD}$ & $\text{A}_{\text{rat}}$ & $\text{SR}^*$ \\
    \midrule
    Power & 33.9  & 32.9  & 32.4  & 31.9  & 31.9  & 31.1  & 30.8  & 30.7  & 30.6  & 30.5  & 30.4  & 30.3  & 29.5  & 29.1  & 28.7  & 28.6  & 28.6  & 28.5  & 28.3  & 27.9 \\
    Gap   & 0.0   & 1.0   & 1.5   & 1.9   & 2.0   & 2.8   & 3.0   & 3.2   & 3.3   & 3.4   & 3.4   & 3.6   & 4.4   & 4.8   & 5.1   & 5.3   & 5.3   & 5.4   & 5.6   & 5.9 \\
    Rank  & 1     & 2     & 3     & 4     & 5     & 6     & 7     & 8     & 9     & 10    & 11    & 12    & 13    & 14    & 15    & 16    & 17    & 18    & 19    & 20 \\
    \midrule
          & $\text{CvM}$ & $\text{AB}_{\text{KL}}$  & $\text{KS}$ & $\text{ZA}$ & $\text{AB}_{\text{Je}}$  & $\text{ZC}$ & $\text{AJ}$ & $\text{AB}_{\text{He}}$  & $\text{KP}$ & $\text{BS}$ & $\text{DLO}_Z$ & $\text{Ho}_V$ & $\text{GV}$ & $\text{Ho}_U$ & $\text{Ho}_K$ & $\text{Ho}_W$ & $\text{AB}_{\chi^2}$  & $\text{Ge}$ & $\text{AB}_{\text{TV}}$  & $\text{SD}$ \\
    \midrule
    Power & 27.9  & 27.8  & 27.3  & 27.3  & 26.6  & 26.3  & 25.1  & 24.9  & 21.2  & 20.8  & 20.7  & 20.7  & 20.6  & 20.2  & 18.8  & 15.8  & 12.5  & 11.9  & 7.0   & 2.8 \\
    Gap   & 5.9   & 6.1   & 6.5   & 6.6   & 7.3   & 7.6   & 8.7   & 9.0   & 12.6  & 13.1  & 13.1  & 13.2  & 13.2  & 13.7  & 15.1  & 18.1  & 21.4  & 22.0  & 26.8  & 31.0 \\
    Rank  & 21    & 22    & 23    & 24    & 25    & 26    & 27    & 28    & 29    & 30    & 31    & 32    & 33    & 34    & 35    & 36    & 37    & 38    & 39    & 40 \\
\end{tabular}%
            \end{center}
            \label{new:table:alpha.0.01.n.20.specialized.ranks}
            \normalsize
        \end{table}
    \end{landscape}
    }


    \newpage
    {
    \pagestyle{empty}
    \newgeometry{left=1.1in,right=1in,top=0.7in,bottom=0.7in,nohead}
    \begin{landscape}
        \def\arraystretch{1.00}
        \begin{table}[htbp]\vspace{-1.51cm}
               \caption{Ordered average \% empirical powers of the 40 tests, for $\alpha=0.01$ and $n=50$, against five groupings of alternative distributions: all models, symmetric, symmetric heavy-tailed, symmetric light-tailed, asymmetric.}
            \begin{center}
                \footnotesize
                \setlength\tabcolsep{2.00pt} 
    \vspace{-0.3cm}
\begin{tabular}{lcccccccccccccccccccc}
    \toprule
    \multicolumn{1}{c}{\textbf{All models}} & $\text{AP}_v$ & $\text{AP}^{(\text{MLE})}_y$ & $\text{AP}_e$ & $\text{AP}_y$ & $\text{ME}^{((1)}_2$ & $\text{ME}^{(2)}_{0.5}$ & $\text{DLO}_X$ & $\text{CK}_v$   & $\text{CK}_c$   & $\text{Wa}$ & $\text{A}_{\text{ent}}$ & $\text{A}_{\text{rat}}$ & $\text{AJ}$ & $\text{LK}$ & $\text{AB}_{\text{He}}$  & $\text{Ku}$ & $\text{AB}_{\text{Je}}$  & $\text{AB}_{\text{KL}}$  & $\text{ZA}$ & $\text{AP}_a$ \\
    \midrule
    Power & 51.1  & 50.6  & 49.9  & 47.9  & 47.4  & 47.0  & 46.4  & 46.2  & 45.6  & 45.4  & 45.0  & 44.2  & 44.0  & 43.0  & 41.9  & 41.3  & 40.7  & 40.3  & 39.9  & 38.3 \\
    Gap   & 0.0   & 0.5   & 1.2   & 3.2   & 3.7   & 4.2   & 4.7   & 4.9   & 5.5   & 5.7   & 6.1   & 6.9   & 7.2   & 8.2   & 9.2   & 9.8   & 10.5  & 10.8  & 11.3  & 12.8 \\
    Rank  & 1     & 2     & 3     & 4     & 5     & 6     & 7     & 8     & 9     & 10    & 11    & 12    & 13    & 14    & 15    & 16    & 17    & 18    & 19    & 20 \\
    \midrule
          & $\text{Z}_K$ & $\text{AD}$ & $\text{DLO}_Z$ & $\text{SR}^*$ & $\text{GV}$ & $\text{Ho}_U$ & $\text{BS}$ & $\text{CvM}$ & $\text{ZC}$ & $\text{KS}$ & $\text{AP}_z$ & $\text{CK}_e$   & $\text{Ho}_V$ & $\text{Ho}_K$ & $\text{AB}_{\text{TV}}$  & $\text{Ho}_W$ & $\text{KP}$ & $\text{Ge}$ & $\text{AB}_{\chi^2}$  & $\text{SD}$ \\
    \midrule
    Power & 38.0  & 37.5  & 37.2  & 37.0  & 37.0  & 36.0  & 36.0  & 35.7  & 35.1  & 35.0  & 33.6  & 33.4  & 31.8  & 29.2  & 29.2  & 24.3  & 23.4  & 15.4  & 14.9  & 5.8 \\
    Gap   & 13.1  & 13.7  & 13.9  & 14.1  & 14.1  & 15.1  & 15.2  & 15.4  & 16.0  & 16.1  & 17.6  & 17.7  & 19.3  & 21.9  & 21.9  & 26.8  & 27.7  & 35.8  & 36.2  & 45.3 \\
    Rank  & 21    & 22    & 23    & 24    & 25    & 26    & 27    & 28    & 29    & 30    & 31    & 32    & 33    & 34    & 35    & 36    & 37    & 38    & 39    & 40 \\
\end{tabular}%
            \end{center}
            \label{new:table:alpha.0.01.n.50.specialized.ranks}
            \normalsize
        \end{table}
    \end{landscape}
    }


    \newpage
    {
    \pagestyle{empty}
    \newgeometry{left=1.1in,right=1in,top=0.7in,bottom=0.7in,nohead}
    \begin{landscape}
        \def\arraystretch{1.00}
        \begin{table}[htbp]\vspace{-1.51cm}
               \caption{Ordered average \% empirical powers of the 40 tests, for $\alpha=0.01$ and $n=100$, against five groupings of alternative distributions: all models, symmetric, symmetric heavy-tailed, symmetric light-tailed, asymmetric.}
            \begin{center}
                \footnotesize
                \setlength\tabcolsep{2.00pt} 
    \vspace{-0.3cm}
\begin{tabular}{lcccccccccccccccccccc}
    \toprule
    \multicolumn{1}{c}{\textbf{All models}} & $\text{DLO}_X$ & $\text{AP}_v$ & $\text{ME}^{(2)}_{0.5}$ & $\text{ME}^{((1)}_2$ & $\text{Wa}$ & $\text{AP}^{(\text{MLE})}_y$ & $\text{LK}$ & $\text{A}_{\text{ent}}$ & $\text{AP}_e$ & $\text{AJ}$ & $\text{A}_{\text{rat}}$ & $\text{AB}_{\text{He}}$  & $\text{CK}_v$   & $\text{Ku}$ & $\text{GV}$ & $\text{DLO}_Z$ & $\text{AB}_{\text{Je}}$  & $\text{AP}_y$ & $\text{ZA}$ & $\text{AB}_{\text{KL}}$ \\
    \midrule
    Power & 62.1  & 60.9  & 60.4  & 60.2  & 59.4  & 59.1  & 58.1  & 56.6  & 56.4  & 56.1  & 55.1  & 55.0  & 53.9  & 53.9  & 53.8  & 53.6  & 53.2  & 52.7  & 52.5  & 51.9 \\
    Gap   & 0.0   & 1.2   & 1.7   & 2.0   & 2.7   & 3.1   & 4.0   & 5.5   & 5.8   & 6.1   & 7.1   & 7.2   & 8.3   & 8.3   & 8.4   & 8.5   & 8.9   & 9.4   & 9.7   & 10.2 \\
    Rank  & 1     & 2     & 3     & 4     & 5     & 6     & 7     & 8     & 9     & 10    & 11    & 12    & 13    & 14    & 15    & 16    & 17    & 18    & 19    & 20 \\
    \midrule
          & $\text{CK}_c$   & $\text{Ho}_U$ & $\text{BS}$ & $\text{AB}_{\text{TV}}$  & $\text{AD}$ & $\text{Z}_K$ & $\text{SR}^*$ & $\text{ZC}$ & $\text{CvM}$ & $\text{KS}$ & $\text{Ho}_V$ & $\text{Ho}_K$ & $\text{AP}_a$ & $\text{CK}_e$   & $\text{Ho}_W$ & $\text{AP}_z$ & $\text{KP}$ & $\text{Ge}$ & $\text{AB}_{\chi^2}$  & $\text{SD}$ \\
    \midrule
    Power & 51.5  & 51.5  & 50.0  & 49.7  & 47.5  & 47.2  & 47.0  & 47.0  & 45.8  & 43.4  & 40.6  & 39.3  & 38.3  & 36.3  & 31.6  & 30.1  & 25.5  & 19.1  & 17.9  & 7.7 \\
    Gap   & 10.6  & 10.7  & 12.2  & 12.4  & 14.6  & 14.9  & 15.1  & 15.1  & 16.3  & 18.8  & 21.5  & 22.9  & 23.8  & 25.8  & 30.6  & 32.0  & 36.7  & 43.1  & 44.3  & 54.5 \\
    Rank  & 21    & 22    & 23    & 24    & 25    & 26    & 27    & 28    & 29    & 30    & 31    & 32    & 33    & 34    & 35    & 36    & 37    & 38    & 39    & 40 \\
\end{tabular}%
            \end{center}
            \label{new:table:alpha.0.01.n.100.specialized.ranks}
            \normalsize
        \end{table}
    \end{landscape}
    }


    \newpage
    {
    \pagestyle{empty}
    \newgeometry{left=1.1in,right=1in,top=0.7in,bottom=0.7in,nohead}
    \begin{landscape}
        \def\arraystretch{1.00}
        \begin{table}[htbp]\vspace{-1.51cm}
               \caption{Ordered average \% empirical powers of the 40 tests, for $\alpha=0.01$ and $n=200$, against five groupings of alternative distributions: all models, symmetric, symmetric heavy-tailed, symmetric light-tailed, asymmetric.}
            \begin{center}
                \footnotesize
                \setlength\tabcolsep{2.00pt} 
    \vspace{-0.3cm}
\begin{tabular}{lcccccccccccccccccccc}
    \toprule
    \multicolumn{1}{c}{\textbf{All models}} & $\text{DLO}_X$ & $\text{Wa}$ & $\text{ME}^{((1)}_2$ & $\text{LK}$ & $\text{ME}^{(2)}_{0.5}$ & $\text{DLO}_Z$ & $\text{AP}_v$ & $\text{Ku}$ & $\text{BS}$ & $\text{AB}_{\text{He}}$  & $\text{GV}$ & $\text{AP}^{(\text{MLE})}_y$ & $\text{AB}_{\text{Je}}$  & $\text{Ho}_U$ & $\text{AB}_{\text{TV}}$  & $\text{AJ}$ & $\text{ZA}$ & $\text{A}_{\text{ent}}$ & $\text{AB}_{\text{KL}}$  & $\text{AP}_e$ \\
    \midrule
    Power & 74.5  & 72.7  & 71.5  & 71.3  & 71.3  & 68.1  & 67.9  & 67.6  & 66.8  & 66.7  & 66.4  & 65.7  & 65.5  & 65.5  & 64.4  & 63.8  & 63.7  & 63.6  & 63.3  & 62.4 \\
    Gap   & 0.0   & 1.8   & 3.0   & 3.1   & 3.1   & 6.4   & 6.6   & 6.9   & 7.7   & 7.8   & 8.0   & 8.7   & 9.0   & 9.0   & 10.1  & 10.7  & 10.8  & 10.9  & 11.1  & 12.1 \\
    Rank  & 1     & 2     & 3     & 4     & 5     & 6     & 7     & 8     & 9     & 10    & 11    & 12    & 13    & 14    & 15    & 16    & 17    & 18    & 19    & 20 \\
    \midrule
          & $\text{ZC}$ & $\text{A}_{\text{rat}}$ & $\text{AP}_y$ & $\text{AD}$ & $\text{CvM}$ & $\text{SR}^*$ & $\text{Z}_K$ & $\text{KS}$ & $\text{CK}_v$   & $\text{CK}_c$   & $\text{Ho}_K$ & $\text{Ho}_V$ & $\text{AP}_a$ & $\text{CK}_e$   & $\text{Ho}_W$ & $\text{AP}_z$ & $\text{KP}$ & $\text{Ge}$ & $\text{AB}_{\chi^2}$  & $\text{SD}$ \\
    \midrule
    Power & 60.2  & 59.2  & 59.2  & 58.7  & 58.5  & 57.7  & 55.5  & 53.7  & 53.4  & 51.2  & 48.2  & 47.6  & 40.2  & 37.8  & 37.1  & 30.3  & 27.7  & 23.2  & 20.8  & 7.7 \\
    Gap   & 14.2  & 15.3  & 15.3  & 15.7  & 16.0  & 16.8  & 18.9  & 20.7  & 21.1  & 23.3  & 26.2  & 26.8  & 34.3  & 36.7  & 37.4  & 44.2  & 46.7  & 51.3  & 53.7  & 66.8 \\
    Rank  & 21    & 22    & 23    & 24    & 25    & 26    & 27    & 28    & 29    & 30    & 31    & 32    & 33    & 34    & 35    & 36    & 37    & 38    & 39    & 40 \\
\end{tabular}%
            \end{center}
            \label{new:table:alpha.0.01.n.200.specialized.ranks}
            \normalsize
        \end{table}
    \end{landscape}
    }


\newpage
   {
    \pagestyle{empty}
    \newgeometry{left=1.1in,right=1in,top=0.7in,bottom=0.7in,nohead}
    \begin{landscape}
        \def\arraystretch{1.00}
        \begin{table}[htbp]\vspace{-1.51cm}
               \caption{The average empirical \% power for each of the 40 tests, for $\alpha=0.10$ and $n=20$, against the 20 specified alternative distributions.\\ The ranges of the free parameter ($k$, $k_1$ or $k_2$) for each distribution are given in Tables~\ref{table:symmetric.alternatives} and~\ref{table:asymmetric.alternatives}. \\
               The highest powers (gap $\le 3.5\%$) are highlighted in grey and the largest is in bold.}

            \begin{center}
                \footnotesize
                \setlength\tabcolsep{2.00pt} 
    \vspace{-0.3cm}
\begin{tabular}{cccccccccccccccccccccc}
    \toprule
    {\bf Type} & {\bf Alternative} &         AD &        CvM &         KS &         Ku &         Wa &        $\text{Z}_K$ &        $\text{Z}_A$ &        $\text{Z}_C$ &      $\text{DLO}_X$ &      $\text{DLO}_Z$ &         Ge &         GV &       $\text{Ho}_K$ &       $\text{Ho}_U$ &       $\text{Ho}_V$ &       $\text{Ho}_W$ &         LK &    $\text{A}_{\text{rat}}$ &  $\text{A}_{\text{ent}}$ &       $\text{AP}_v$ \\
    \midrule
     \multirow{6}{*}{Sym., heavy} & GED($k<1$)   &  44.9 & 45 & 43.2 & \cellcolor{gray!25} 48.1 & \cellcolor{gray!25} 49.4 & 43.2 & 45 & 45 & \cellcolor{gray!25} {\bf 50.3} & \cellcolor{gray!25} 49.1 & \cellcolor{gray!25} 48.9 & 41.6 & 34.4 & \cellcolor{gray!25} 47.1 & 42 & 24.1 & 45.4 & 21.1 & 24.9 & 31.9 \\
      & $t(k<3.5)$   &  36.2 & 34.4 & 33.5 & 37.2 & 38.1 & 37.6 & 41.1 & 41.9 & 41.7 & 41.1 & \cellcolor{gray!25} {\bf 45.7} & 41 & 39.5 & \cellcolor{gray!25} 43.3 & 40.5 & 20.6 & 39.4 & 23.8 & 27.2 & 32.9 \\
      & MixL($w=0.5,k_1=0,k_2>1$)   &  80.5 & 83.9 & 81.8 & \cellcolor{gray!25} 87.6 & \cellcolor{gray!25} {\bf 88.5} & 73.7 & 69 & 64.7 & 83.9 & 81.4 & 61.4 & 52 & 28.1 & 69.9 & 53.4 & 28.6 & 67.4 & 28.7 & 40.1 & 51.4 \\
      & MixN($w=0.5,k_1=0,k_2>3.5$)   &  74 & 79.5 & 78.1 & \cellcolor{gray!25} 85 & \cellcolor{gray!25} {\bf 85.8} & 66.3 & 52.5 & 46.7 & 71.3 & 69.3 & 35.6 & 35.2 & 10.5 & 48 & 31.3 & 16.6 & 51.4 & 20.9 & 28.2 & 36.7 \\
      & NIG($k_1<0.6,k_2=0$)   &  20.3 & 18.7 & 18.2 & 21.1 & 22 & 21.8 & 25.2 & 26 & 25.5 & 24.7 & \cellcolor{gray!25} {\bf 30.7} & 24.9 & 23.3 & 26.6 & 24.4 & 17.1 & 23.3 & 11 & 12.3 & 16.4 \\
      & Tu($k<-0.14$)   &  54.9 & 53.1 & 51.2 & 56.8 & 58.2 & 55.4 & 59.2 & 59.8 & \cellcolor{gray!25} 61.6 & 60.7 & \cellcolor{gray!25} {\bf 64.5} & 57.8 & 53.5 & \cellcolor{gray!25} 62 & 57.7 & 26.8 & 58.7 & 31.8 & 37.2 & 46.2 \\
\bottomrule
    \end{tabular}%
            \end{center}
            \label{new:table:alpha.0.10.n.20}
            \normalsize
        \end{table}
    \end{landscape}
    }


        \newpage
   {
    \pagestyle{empty}
    \newgeometry{left=1.1in,right=1in,top=0.7in,bottom=0.7in,nohead}
    \begin{landscape}
        \def\arraystretch{1.00}
        \begin{table}[htbp]\vspace{-1.51cm}
               \caption{The average empirical \% power for each of the 40 tests, for $\alpha=0.10$ and $n=50$, against the 20 specified alternative distributions.\\ The ranges of the free parameter ($k$, $k_1$ or $k_2$) for each distribution are given in Tables~\ref{table:symmetric.alternatives} and~\ref{table:asymmetric.alternatives}. \\
               The highest powers (gap $\le 3.5\%$) are highlighted in grey and the largest is in bold.}

            \begin{center}
                \footnotesize
                \setlength\tabcolsep{2.00pt} 
    \vspace{-0.3cm}
%
            \end{center}
            \label{new:table:alpha.0.10.n.50}
            \normalsize
        \end{table}
    \end{landscape}
    }



        \newpage
   {
    \pagestyle{empty}
    \newgeometry{left=1.1in,right=1in,top=0.7in,bottom=0.7in,nohead}
    \begin{landscape}
        \def\arraystretch{1.00}
        \begin{table}[htbp]\vspace{-1.51cm}
               \caption{The average empirical \% power for each of the 40 tests, for $\alpha=0.10$ and $n=100$, against the 20 specified alternative distributions.\\ The ranges of the free parameter ($k$, $k_1$ or $k_2$) for each distribution are given in Tables~\ref{table:symmetric.alternatives} and~\ref{table:asymmetric.alternatives}. \\
               The highest powers (gap $\le 3.5\%$) are highlighted in grey and the largest is in bold.}

            \begin{center}
                \footnotesize
                \setlength\tabcolsep{2.00pt} 
    \vspace{-0.3cm}
%
            \end{center}
            \label{new:table:alpha.0.10.n.100}
            \normalsize
        \end{table}
    \end{landscape}
    }



        \newpage
   {
    \pagestyle{empty}
    \newgeometry{left=1.1in,right=1in,top=0.7in,bottom=0.7in,nohead}
    \begin{landscape}
        \def\arraystretch{1.00}
        \begin{table}[htbp]\vspace{-1.51cm}
               \caption{The average empirical \% power for each of the 40 tests, for $\alpha=0.10$ and $n=200$, against the 20 specified alternative distributions.\\ The ranges of the free parameter ($k$, $k_1$ or $k_2$) for each distribution are given in Tables~\ref{table:symmetric.alternatives} and~\ref{table:asymmetric.alternatives}. \\
               The highest powers (gap $\le 3.5\%$) are highlighted in grey and the largest is in bold.}

            \begin{center}
                \footnotesize
                \setlength\tabcolsep{2.00pt} 
    \vspace{-0.3cm}
\begin{tabular}{cccccccccccccccccccccc}
    \toprule
    {\bf Type} & {\bf Alternative} &         AD &        CvM &         KS &         Ku &         Wa &        $\text{Z}_K$ &        $\text{Z}_A$ &        $\text{Z}_C$ &      $\text{DLO}_X$ &      $\text{DLO}_Z$ &         Ge &         GV &       $\text{Ho}_K$ &       $\text{Ho}_U$ &       $\text{Ho}_V$ &       $\text{Ho}_W$ &         LK &    $\text{A}_{\text{rat}}$ &  $\text{A}_{\text{ent}}$ &       $\text{AP}_v$ \\
    \midrule
     \multirow{6}{*}{Sym., heavy} & GED($k<1$)   &  73.9 & 74.7 & 73 & 77.6 & \cellcolor{gray!25} 79 & 72.2 & 71.7 & 73.3 & \cellcolor{gray!25} 80.4 & \cellcolor{gray!25} {\bf 81.7} & 76.9 & \cellcolor{gray!25} 79.4 & 65.6 & \cellcolor{gray!25} 79.7 & 68.9 & 59.8 & 78.1 & 54.4 & 46.7 & 76.6 \\
      & $t(k<3.5)$   &  66.8 & 63.3 & 62.2 & 68.8 & 70.6 & 75.6 & 77.8 & 80.2 & 72.5 & 73.3 & \cellcolor{gray!25} {\bf 83.8} & 79.5 & \cellcolor{gray!25} 80.7 & 79.5 & 79.4 & 77.3 & 72.1 & 39.2 & 61.8 & 78.7 \\
      & MixL($w=0.5,k_1=0,k_2>3.5$)   &  78.6 & 78.8 & 76.8 & 81.7 & \cellcolor{gray!25} 82.9 & 76.3 & 75.8 & 77 & \cellcolor{gray!25} 84.4 & \cellcolor{gray!25} {\bf 85.6} & 78.7 & \cellcolor{gray!25} 84.1 & 45.5 & \cellcolor{gray!25} 84.4 & 59 & 32.4 & \cellcolor{gray!25} 82.8 & 56.3 & 27.9 & 80.8 \\
      & MixN($w=0.5,k_1=0,k_2>3.5$)   &  84.4 & 84.4 & 82.4 & \cellcolor{gray!25} 89.4 & \cellcolor{gray!25} {\bf 90.8} & 73.7 & 58 & 59 & 80.7 & 84.1 & 18 & 65.8 & 1.2 & 67.1 & 4.4 & 5.6 & 79.5 & 80 & 1.9 & \cellcolor{gray!25} 88.3 \\
      & NIG($k_1<0.65,k_2=0$)   &  44.2 & 41.6 & 40.8 & 47.7 & 49.6 & 51.4 & 52 & 55.8 & 52.6 & 54 & \cellcolor{gray!25} {\bf 63.8} & 59.9 & 53.3 & 60 & 53.5 & 47.3 & 51.8 & 15.6 & 25.6 & 54.9 \\
\bottomrule
    \end{tabular}%
            \end{center}
            \label{new:table:alpha.0.10.n.200}
            \normalsize
        \end{table}
    \end{landscape}
    }


    \newpage
    {
    \pagestyle{empty}
    \newgeometry{left=1.1in,right=1in,top=0.7in,bottom=0.7in,nohead}
    \begin{landscape}
        \def\arraystretch{1.00}
        \begin{table}[htbp]\vspace{-1.51cm}
               \caption{Ordered average \% empirical powers of the 40 tests, for $\alpha=0.10$ and $n=20$, against five groupings of alternative distributions: all models, symmetric, symmetric heavy-tailed, symmetric light-tailed, asymmetric.}
            \begin{center}
                \footnotesize
                \setlength\tabcolsep{2.00pt} 
    \vspace{-0.3cm}
%
            \end{center}
            \label{new:table:alpha.0.10.n.20.specialized.ranks}
            \normalsize
        \end{table}
    \end{landscape}
    }


    \newpage
    {
    \pagestyle{empty}
    \newgeometry{left=1.1in,right=1in,top=0.7in,bottom=0.7in,nohead}
    \begin{landscape}
        \def\arraystretch{1.00}
        \begin{table}[htbp]\vspace{-1.51cm}
               \caption{Ordered average \% empirical powers of the 40 tests, for $\alpha=0.10$ and $n=50$, against five groupings of alternative distributions: all models, symmetric, symmetric heavy-tailed, symmetric light-tailed, asymmetric.}
            \begin{center}
                \footnotesize
                \setlength\tabcolsep{2.00pt} 
    \vspace{-0.3cm}
%
            \end{center}
            \label{new:table:alpha.0.10.n.50.specialized.ranks}
            \normalsize
        \end{table}
    \end{landscape}
    }


    \newpage
    {
    \pagestyle{empty}
    \newgeometry{left=1.1in,right=1in,top=0.7in,bottom=0.7in,nohead}
    \begin{landscape}
        \def\arraystretch{1.00}
        \begin{table}[htbp]\vspace{-1.51cm}
               \caption{Ordered average \% empirical powers of the 40 tests, for $\alpha=0.10$ and $n=100$, against five groupings of alternative distributions: all models, symmetric, symmetric heavy-tailed, symmetric light-tailed, asymmetric.}
            \begin{center}
                \footnotesize
                \setlength\tabcolsep{2.00pt} 
    \vspace{-0.3cm}
%
            \end{center}
            \label{new:table:alpha.0.10.n.100.specialized.ranks}
            \normalsize
        \end{table}
    \end{landscape}
    }


    \newpage
    {
    \pagestyle{empty}
    \newgeometry{left=1.1in,right=1in,top=0.7in,bottom=0.7in,nohead}
    \begin{landscape}
        \def\arraystretch{1.00}
        \begin{table}[htbp]\vspace{-1.51cm}
               \caption{Ordered average \% empirical powers of the 40 tests, for $\alpha=0.10$ and $n=200$, against five groupings of alternative distributions: all models, symmetric, symmetric heavy-tailed, symmetric light-tailed, asymmetric.}
            \begin{center}
                \footnotesize
                \setlength\tabcolsep{2.00pt} 
    \vspace{-0.3cm}
%
            \end{center}
            \label{new:table:alpha.0.10.n.200.specialized.ranks}
            \normalsize
        \end{table}
    \end{landscape}
    }

\end{document}